\documentclass[fleqn,usenatbib]{mnras}

\usepackage[T1]{fontenc}
\usepackage{ae,aecompl}
\usepackage{graphicx}	
\usepackage{amsmath}	
\usepackage{amssymb}	
\usepackage{longtable}
\usepackage{multirow}
\usepackage{verbatim}

\newcommand{\ibh}{{\rm IMBH}}
\newcommand{\bh}{{\rm BH}}
\newcommand{\bhs}{{\rm BHS}}
\newcommand{\gc}{{\rm GC}}
\newcommand{\oth}{{\rm oth}}
\newcommand{\Ms}{~{\rm M}_\odot}
\newcommand{\df}{{\rm df}}

\newcommand{\Log}{{\rm Log}}

\title[BHs subsystems in globular clusters]{MOCCA SURVEY DATABASE I. Unravelling black hole subsystems in globular clusters}

\author[M. Arca Sedda, A. Askar and M. Giersz]{Manuel Arca Sedda$^{1}$\thanks{E-mail:m.arcasedda@ari.uni-heidelberg.de}, Abbas Askar$^{2}$, Mirek Giersz$^{2}$
\\
$^{1}$Astronomisches Rechen-Institut, Zentrum f{\"u}r Astronomie, University of Heidelberg, M{\"o}nchhofstrasse 12-14, 69120, Heidelberg, Germany\\
$^{2}$Nicolaus Copernicus Astronomical Center, Polish Academy of Sciences, 
ul. Bartycka 18, 00-716 Warsaw, Poland\\
}

\date{Accepted XXX. Received YYY; in original form ZZZ}
\pubyear{2017}

\begin{document}
\label{firstpage}
\pagerange{\pageref{firstpage}--\pageref{lastpage}}
\maketitle

\begin{abstract}
In this paper, we discuss how globular clusters (GCs) structural and observational properties can be used to infer the presence of a black hole system (BHS) inhabiting their inner regions.
We propose a novel way to identify the BHS size, 
defined as the GC radius containing a mass contributed equally from stars and stellar BHs.
Using this definition, similar to the well-known concept of ``influence radius'', we found a ``fundamental plane'' connecting the BHS typical density with the GC central surface density profile, total luminosity and observational half-mass radius. Our approach allows us to define a unique way to connect the observational GCs parameters with their dark content. Comparing our results with observed Milky Way GCs, we found that many of them likely host, at the present time, as many as several hundreds of BHs. These BHS are characterized by a relatively low typical density, $\rho_\bhs \sim 10-10^5\Ms$ pc$^{-3}$ and composed of relatively massive BHs,
with average masses in the range $m_\bhs = 14-22\Ms$. 
We also show that a similar approach can be used to find Milky Way GCs potentially hosting an intermediate-mass black hole. 
\end{abstract}

\begin{keywords}
black hole physics,
(Galaxy:) globular clusters: general,
stars: black holes,
stars: kinematics and dynamics
\end{keywords}

\section{Introduction}

The formation of stellar black holes (BH), representing the ultimate stage of dying stars with an initial mass $\gtrsim 15-20 \Ms$, is a process occurring on time-scales of a few to tens of Myr. Tens to thousands of such BHs are expected to form in dense stellar environments, such as globular (GCs) or nuclear clusters (NCs) \citep{kulkarni93,sigurdsson93}. At their birth, BHs may receive strong natal kicks that potentially can lead to their ejection from the parent cluster. However, the recoiling velocity amplitude is still a matter of debate.
A number of works suggested that BHs the kick distribution is similar to what expected for neutron stars \citep{repetto12,janka13,sippel13,mandel16}. However, it seems possible that massive stars can undergo direct collapse, turning into BHs without losing a large mass fraction and avoiding supernova explosion \citep{adams2017}. According to this picture, BHs might have masses quite larger than previously thought, and experience no or low natal kicks
\citep{fryer99,belckzinski10,fryer12,spera15}. 

Small recoiling velocities imply that the BHs retention fraction in dense stellar clusters is much larger than previously thought. A larger retention fraction is also required to explain the ever-growing observational evidence of BHs signatures in Galactic GCs \citep{strader12,chomiuk13,miller15,bahramian17,giesers18}, whose presence seem largely supported by recent numerical works \citep{morscher13,wong14,morscher15,repetto15,peuten16}. 

Retained BHs would undergo rapid mass segregation, populating the inner regions of their parent cluster and likely forming a subsystem on a core-collapse time-scale \citep{spitzer87,portegies00,zwart02,fregeau2004,zwart04,gaburov08,freitag06c,AS16}.

Three-body interactions and multiple scatterings can drive the formation of BH binaries, which act as a power supply for the cluster core. These binaries kick out the most massive BHs, depleting the global BH reservoir, and eventually kick out each other through super-elastic encounters \citep{banerjee10,downing10,rodriguez15,askar17}. 
The single-binary and binary-binary continuous interactions lead BH binaries to get harder and harder, until they are ejected from the cluster core or merge in there releasing gravitational waves (GWs) \citep{portegies00,banerjee10,downing10,rodriguez15,rodriguez16,wang16,askar17}. 
Stars interacting with retained BHs are pushed on wider orbits, causing the expansion of the GC core and delaying core-collapse \citep{merritt04,mackey08,gieles10,wang16}.

Currently, there is no general consensus on the definition of a BHS. Recently, \cite{breen13} investigated how a population of BHs behaves in idealized star cluster models, revisiting the pioneering work made by \cite{spitzer71} \cite[but see also][]{kulkarni93,sigurdsson93}. 
In the case of a so-called ``Spitzer-instable'' system, the BHs lose energy to the other stars and segregate toward the GC centre, causing a progressive reduction of the BHS half-mass radius. This contraction slows down as soon as the BHS energy is transferred to the surrounding stars and is balanced by the ``thermal energy'' provided by the BHs binaries formed in the very inner portion of the GC. The complex interactions provide sufficient energy to avoid further collapses, although a stable configuration is hardly achievable \citep{spera16,bianchini16}.

Under the simple assumption of a two-mass population of objects and using theoretical arguments, \cite{breen13} have shown that the energy generated from the repeated scattering between the light and heavy objects, which flows through the GC half-mass radius, regulates the evolution of the heavier component, which settles into the GC centre. 
Their results suggest that GCs having a sufficiently long half-mass relaxation time can retain a sizeable number of BHs. However, is quite hard to define a BHS since BHs are typically mixed with other stars in realistic GC models, making difficult to define the BHS size and structure. 

In this paper, we propose a novel method to define the BHS radial extent. Our definition of BHS radius is similar to the ``influence radius'' defined for supermassive BHs that inhabit galactic nuclei \citep{peebles72,merritt06,merritt13}.

Using a large suite of GC models simulated in the context of the ``MOCCA-Survey Database I project'', we determined a set of scaling relations aimed at allowing us to infer the presence of a BHS, and its main properties, through the observational and structural features of its host cluster. We found that GCs having a BHS are distributed in a narrow region of the surface brightness - average surface luminosity plane, well detached from GCs having a central IMBH or that exhibit none of these ``dark'' features.
These relations represent a unique tool to unveil the presence of a BHS in the centre of GCs. In a companion paper, we use these correlations to select a sample of 29 Galactic GCs that may harbor a BHS in their centre \citep{askar18}.

The paper is organized as follows: in section \ref{label:gcm} we briefly describe the MOCCA numerical models used in this work; in section \ref{sec:bhsingcs} we introduce our definition of BHS, discussing the basic relations connecting the BHS main properties; section \ref{label:scali} is focused on the scaling relations connecting the BHS main parameters and the host cluster structural properties; section \ref{sec:obser} presents the fundamental relation that allows connecting the GC observational properties to the BHS density. Finally, in section \ref{sec:end} are summarised the conclusions of this work.

\section{Globular cluster models}
\label{label:gcm}

\subsection{The MOCCA SURVEY DATABASE I.}
In this paper, we use the results from the MOCCA-Survey Database I \citep{askar17} that comprises of about 2000 realizations of GCs with different initial masses, structural and orbital parameters. These models were simulated with the MOCCA code for star cluster simulations, which treats the relaxation process using the method described by \citet{henon71}, conveniently improved by \citet{stdo86}, and recently by \cite{giersz08} \citep[but see also ][ and reference therein]{giersz13,Giersz15}. MOCCA implements the SSE and BSE codes \citep{hurley00,hurley2002} for treating binary and stellar evolution, while strong binary-single and binary-binary interactions are handled by the \texttt{FEWBODY} code \citep{fregeau2004}. 
The initial parameters of the models simulated in the MOCCA-Survey Database I can be found in Table 1 in \citep{askar17}.
In nearly half of the simulated models,  supernovae natal kick velocities for neutron stars and BHs are assigned according to a Maxwellian distribution, assuming a dispersion of $265$ km s$^{-1}$ \citep{hobbs}. In the remaining cases,  BH natal kicks were modified according to the mass fallback procedure described by \cite{belczynski02}. The model metallicities are selected between  $Z = 0.0002, ~0.001, ~0.005, ~0.006$ or $0.02$.

All MOCCA models are characterized by a \cite{kroupa01} initial mass function, with a minimum and maximum initial stellar mass of $0.08\Ms$ and $100 \Ms$, respectively.
The total number of objects sampled in our simulated GCs are $4\times 10^4$, $10^5$, $4\times 10^5$, $7\times 10^5$ and $1.2\times 10^6$, including both single stars and primordial binaries. All our GCs are described by \cite{King} models, with central concentration parameters values $W_0 = 3,~6$ and $9$. We assumed an initial tidal radius $R_t = 30,~60$ or $120$ pc, while the ratio between the tidal radius and the GC half-mass radius is $50$, $25$ or the model is tidally-filling. We allowed for four different values of the primordial binary fraction: $5\%$, 
$10\%$, $30\%$ and $95\%$. In models characterised by an initial binary fraction equal to or lower than $30\%$, we selected the initial eccentricities of binary systems according to a thermal distribution \citep{jeans19}, the semi-major axes according to a flat logarithmic distribution, and the mass ratio according to a flat distribution. For models containing a larger binary fraction, instead, the initial binary properties are selected according to the distribution provided by \citet{kroupa95,kroupa11}.

The GCs are assumed to move on a circular orbit at Galactocentric distances between $1$ and $50$ kpc. The Galactic potential is modelled in the simple point-mass approximation, taking as central mass the value of the Galaxy mass enclosed within the GCs orbital radius. 

As pointed out in \citet{askar17}, the initial conditions assumed to create the MOCCA-Survey Database I were not specifically selected to reproduce the Galactic GC population. Nevertheless, their observational parameters calculated at the present-day exhibit a remarkably good agreement with Milky Way GCs.

\subsection{Globular clusters hosting a black hole subsystem}

In order to focus on the BHS properties, we selected models retaining at least 10 BHs after 12 Gyr. Our subsample, comprised of $N_{\rm sub}=172$ out of the over 2000 simulated systems,  contains GCs with different properties, spanning a wide range of initial masses, binary fraction, and initial metallicity.
\\
The corresponding initial mass distribution peaks at $M_\gc \sim 6.3\times 10^5\Ms$, with $\sim 154$ models having masses in between $M_\gc\sim 3.2-10\times 10^5\Ms$ while the remaining are smaller ($M_\gc \lesssim 2.2\times 10^5\Ms$).
The GC inital core radii ($r_c$) are smaller than 1.5 pc in nearly $80\%$ of the models, with more than 50 models having $r_c = 0.8$ pc. Nearly $90$ models have an initial half-mass radius ($r_h$) smaller than 3 pc, 55 have $r_h = 4.5$ pc while the remaining are more extended tidally-filling models having $6<r_h<17$ pc.
The initial \cite{King} $W_0$ parameter is evenly distributed among small $W_0 = 3$ (65 models), intermediate $W_0 = 6$ (70 models), while a smaller number of cases have higher values $W_0 = 9$ (37 models).
The GCs initial relaxation time, $T_{\rm rel}$, varies in a wide range: $\lesssim 0.65 Gyr$ (19 models), $\simeq 1-2 Gyr$ (118 models), $\gtrsim 10 Gyr$ (35 models).

More than a half of the GCs in the sample (96) are characterised by metallicities around $ Z = 10^{-3}$, while a few models have $Z \lesssim 2.5\times 10^{-4}$ (9), and 42 models have sub-solar metallicities ( $Z \sim 6\times 10^{-3} $). In the remaining 25 models, instead, the initial metallicity assumes solar-values.
In all the models containing a BHS, the BH natal kicks was calculated taking into account the amount of matter that fallbacks after supernova explosion, according to \cite{belczynski02}.

Our sample consists of GCs having a low initial binary fraction ($f_{\rm bin}\leq 0.1$ for 88 models) or intermediate values ($0.1 < f_{\rm bin} \leq 0.3$ for 20 models), while a substantial fraction is ``binary-rich'' ($f_{\rm bin}=0.95$ for 64 models). 

Overall, the sample seems quite heterogeneous and is characterised by quite different initial conditions, thus highlighting at a glance that BHS can be a common feature of GCs. In the next section we will show how it is possible to infer the BHS main parameters from the observational and structural properties of the parent cluster.

\begin{figure}
\centering
\includegraphics[width=8cm]{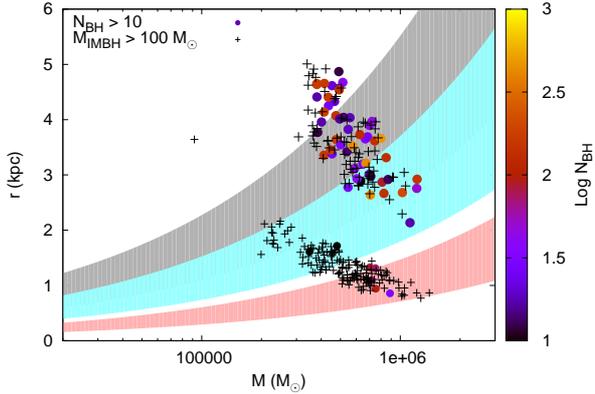}
\caption{ 
MOCCA GCs initial Galactocentric radius $R$ (Y axis) and initial total mass $M$ (X axis).The color-coded map marks the retained number of BHs after 12 Gyr. The shaded regions identify $M-R$ couples characterized by $t_\df = 1$ (red region), $5$ (cyan region) and $12$ Gyr (grey region). The lower boundary of each region represents $t_\df$ for circular orbits, while the upper boundary marks the limit in which the GC moves on a nearly radial orbit. Filled circles represent GCs having at least 10 BHs at 12 Gyr, while crosses identify those hosting an IMBH.
}
\label{F1}
\end{figure}

Interestingly, a substantial number of our selected models ($\sim 120$) have large galactocentric radii, $R_0 > 2$ kpc, although almost $1/3$ of them
orbits at smaller distances from the Galactic Centre. GCs moving at smaller distances have masses in between $4\times 10^5 \Ms$ and $1.1\times 10^6\Ms$.
Clusters having sufficiently small apocentres can segregate toward the Galactic Centre due to the intense action of dynamical friction (df) \citep{Trem76,Dolc93}. 
Figure \ref{F1} shows how the df time-scale $t_\df$ varies at varying GCs masses and Galactocentric distances for the MOCCA models containing either an intermediate mass BH (IMBH) with mass above $10^2\Ms$ or at least 10 BHs after 12 Gyr. 
The df time is calculated following \cite{ASCD14a} (but see also \cite{ASCD15He}), according to which $t_{\rm df}\propto M_\gc^{-0.67}R_0^{1.76}$.

To represent the Milky Way we used the model recently provided by \cite{kafle14}, consisting of a \cite{Her90} sphere with scale length $\sim 11$ kpc and total mass $6\times 10^{11}\Ms$.

We see that a substantial number of MOCCA models with large masses and small Galactocentric distances will quickly diffuse to the Galactic center reducing substantially the number of models with (IMBH), but not strongly influencing the number of models with BHSs. To form an IMBH, clusters have to be initially very dense \citep{Giersz15},- massive with small tidal radius,  but to sustain BHSs until the Hubble time clusters need to be initially not too dense \citep{breen13} - relatively large half-mass radius.

In order to provide a very preliminary investigation about whether the GCs global properties can be used also to infer the presence of an IMBH in their inner regions, we selected 470 MOCCA models hosting a central BH heavier than $150\Ms$ at 12 Gyr. We stress here that this subject will be deeply discussed in a companion paper.

Note that the possibility that some GCs deliver their IMBH or BHS toward the galactic centre can have interesting implications for IMBH-SMBH pairing and coalescence events, as recently investigated by \citep[Arca Sedda and Gualandris, in prep.]{ASCD17b,fragione17}.

Orbitally segregated GCs can deposit into the hosting galactic centre a substantial population of BHs living in binary systems. For instance, the progenitor of low-mass X-ray binaries (LMXBs), containing either an NS or a stellar BH, can easily be transported into the galactic centre from inspiral clusters.
As a consequence, the population of LMXBs inhabiting the galactic inner regions might benefit from the GC infall process. 

Recently, detailed observations of the Milky Way nuclear cluster (MWNC) revealed the presence of as many as 20000 BHs probably orbiting the SMBH surroundings \citep{hailey18}. As suggested by a number of works, most of the MWNC likely formed through repeated mergers of $\sim 10-20$ massive star clusters with masses above $\simeq 10^6\Ms$ \citep{AMB,antonini14,ASCD14b,ASCD15He,ASK17}. 
As we will show in detail in the following, BHS constitute nearly the $70\%$ of the GC total BH reservoir. Assuming that one BH form every 1000 stars, which is expected from standard stellar evolution, this means that infalling clusters might have brought to the MWNC $\sim 0.7\times 10^{-3} \times 10^6 \times 20 = 15000$ BHs, either as a single object or in a binary system. This number fits nicely with the values inferred recently by \citep{hailey18}.  
In a subsequent paper, we will explore whether delivered BHs can lead to the formation of a number of LMXB containing a stellar BH consistent with the latest observations and modelling \citep{generozov18}.

\section{Black Hole Subsystems in globular clusters}
\label{sec:bhsingcs}

\subsection{A novel definition for BH subsystem}
As shown by \cite{breen13}, in the idealized case that a massive GC can be modeled as a two-mass population system, the energy exchange rate between the BHS and the surrounding stars depend on the energy flow through the GC half-mass radius $r_h$ and the corresponding half-mass relaxation time $t_{rh}$. 
In particular, they suggest that the ratio between the BHS and the GC core radius scale as the ratio between the average BHS and GC mass and the ratio of their total masses
\begin{equation}
\frac{r_{{\rm BHS},h}}{r_{{\rm GC},h}} \propto \left(\frac{m_\bhs}{m_\gc}\right)^{2/5}\left(\frac{M_\bhs}{M_\gc}\right)^{3/5}.
\end{equation}
This implies that to sustain a BHS up to the Hubble time, the GC half-mass relaxation time has to be larger than about 1 Gyr \citep{breen13}.

We note here that, as long as this relation remains valid, it can have profound implications on the BHS lifetime. 
The most massive BHs will be ejected in strong binary-binary and binary-single encounters Promptly after the BHS core-collapse, thus reducing the total BHS mass and its average mass as well.  As a consequence, the outward flux energy generated by the BH-BH/BH-stars interactions decreases and the BHS contracts. This, in turn, drives a density increase and a consequent enhancement of the dynamical interactions rate until they can sustain the energy flow. 

Hence, GCs having a large initial relaxation time should contain massive and extended BHS. 

However, as long as new binaries form and multi-body processes occur efficiently, resulting in the depletion of BHs, the energy supply is insufficient and the BHS slowly dissolves into the sea of other stars.

During these complex stages, which last on time-scales comparable to the half-mass relaxation times, the BHS can be sufficiently dense to mimic the effect of an intermediate-mass black hole, exhibiting similar scaling relations with the host GC mass \cite{AS16}.

Since BHs are usually ``mixed'' with other stars, a natural definition of BHS radius would be the region where BHs play a dominant role in determining the dynamics. Following this idea, we define the BHS size as the sphere enclosing 
$50\%$ of the cumulative mass in BHs and the remaining in other stars.
By definition, the radius of this sphere, $R_\bhs$, encloses
twice the total mass of the BHS, thus representing an analogous of the well-known ``influence radius'' calculated for an isothermal sphere \citep{merritt13}. In fact, $R_\bhs$ defined this way marks the region over which BHs affect significantly the host GC inner dynamics.

\begin{figure}
\centering\includegraphics[width=8cm]{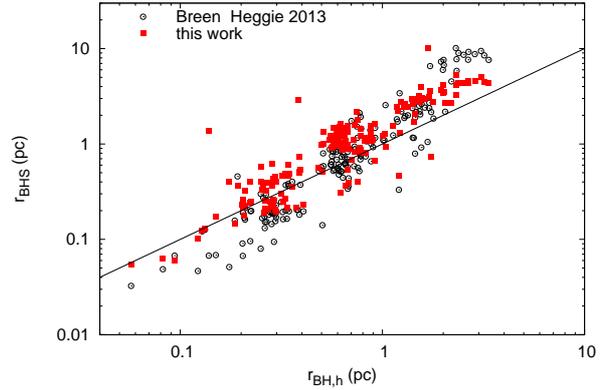}
\caption{Our definition of BHS radius (filled red squares) and \citet{breen13} predicted values (open black circles) as a function of the actual BH half-mass radius as calculated for our MOCCA sample. The straight black line represents the equality between calculated and predicted values, i.e. $f(x) = x$.}
\label{F2}
\end{figure}

To investigate possible similarities between our definition of BHS size and \cite{breen13} theoretical predictions, we show in Fig. \ref{F2} how do they compare with the actual BH half-mass radius, calculated at 12 Gyr for all the MOCCA models hosting more than 10 BHs.

\cite{breen13} definition of BHS size seems to over-predict the actual BH half-mass radius, especially for values above 1 pc. Interestingly, our definition agrees pretty well with the real $r_{\bh{\rm,h}}$ value.

A rough explanation for the similarity between $r_{\bh{\rm,h}}$ and $R_\bhs$ can be developed following simple arguments. Let's assume that the BH mass distribution can be described by an isothermal sphere, 
\begin{equation}
M_\bh(r) = \frac{\sigma_\bh^2}{(2\pi G)} r,
\end{equation}
being $\sigma_\bh$ the central velocity dispersion of the BHs population. Therefore, the resulting BH half-mass radius will be given by:
\begin{equation}
r_{\bh{\rm, h}} = \frac{\pi G}{\sigma_\bh^2} M_\bh .
\end{equation}

The corresponding GC mass enclosed within $r_{\bh,{\rm h}}$ can be calculated as
\begin{equation}
M(r_{\bh,{\rm h}}) = \frac{1}{2}\left(\frac{\sigma}{\sigma_\bh}\right)^2 M_\bh.
\end{equation}

Hence, under the hypothesis of equilibrium between stars and BHs, $\sigma_\bh \sim \sigma$,the BH half-mass radius contains the same amount of mass in stars and BHs, and this roughly corresponds to $50\%$ of the whole BH mass.

This is clearly an oversimplification of the whole picture, but provides a simple explanation for the similarity between the BH half-mass radius and our definition of BHS size. As we will discuss in the next section, the BHS defined here can contain up to $70\%$ of the BHs total mass, thus deviating from the half-mass radius. However, we will show that using $R_\bhs$ instead of $r_{\bh{\rm ,h}}$ allows us to provide a large set of tight scaling relations connecting the GC and the BHs properties.

\subsection{BHS basic properties}
\label{label:scali}

Following the aforementioned assumptions, we define the BHS mass ($M_\bhs$) as the mass in BHs enclosed within $R_\bhs$, while $N_\bhs$ is the number of BHs inside $R_\bhs$, $m_\bhs = M_\bhs / N_\bhs$ represents the BHS average mass\footnote{Note that this is the mean mass of BHs contained within $R_\bhs$.},  and $\rho_\bhs = M_\bhs/R_\bhs^3$ the BHS typical density.

\begin{figure*}
\includegraphics[width=8cm]{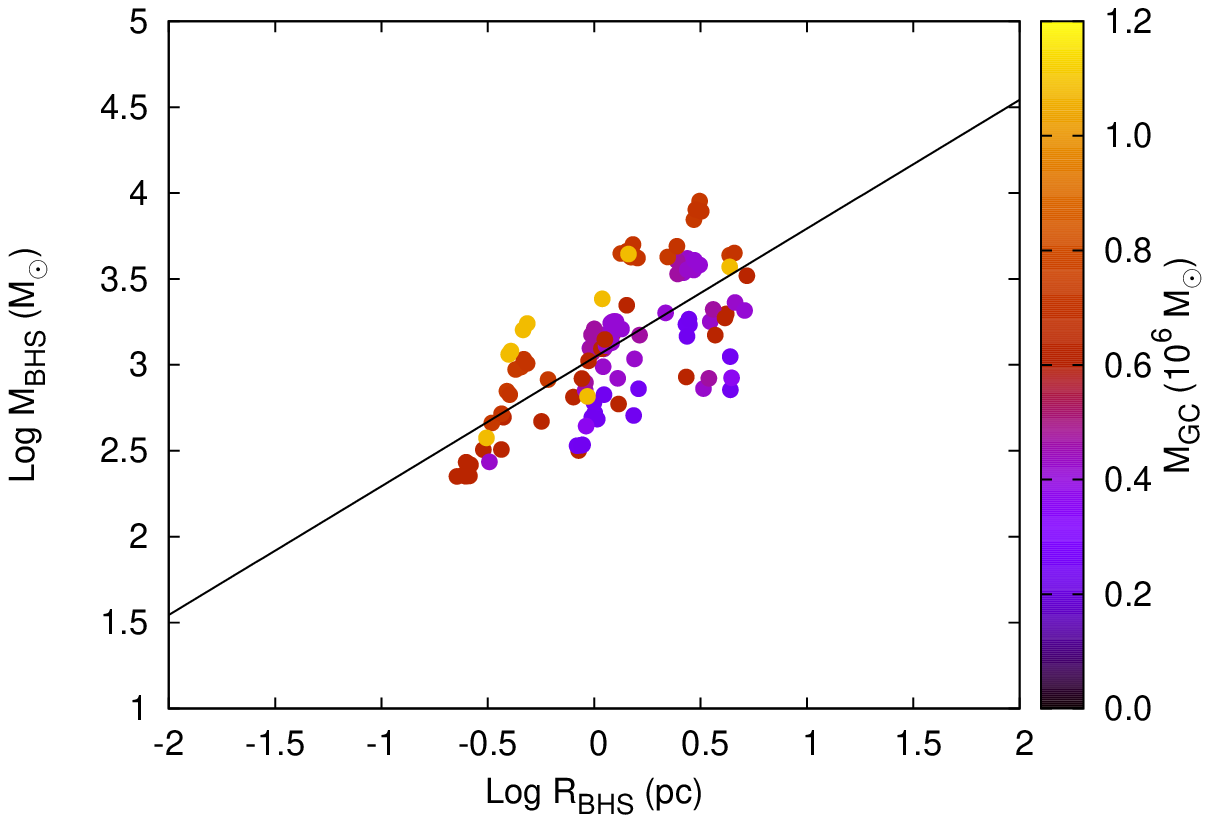}
\includegraphics[width=8cm]{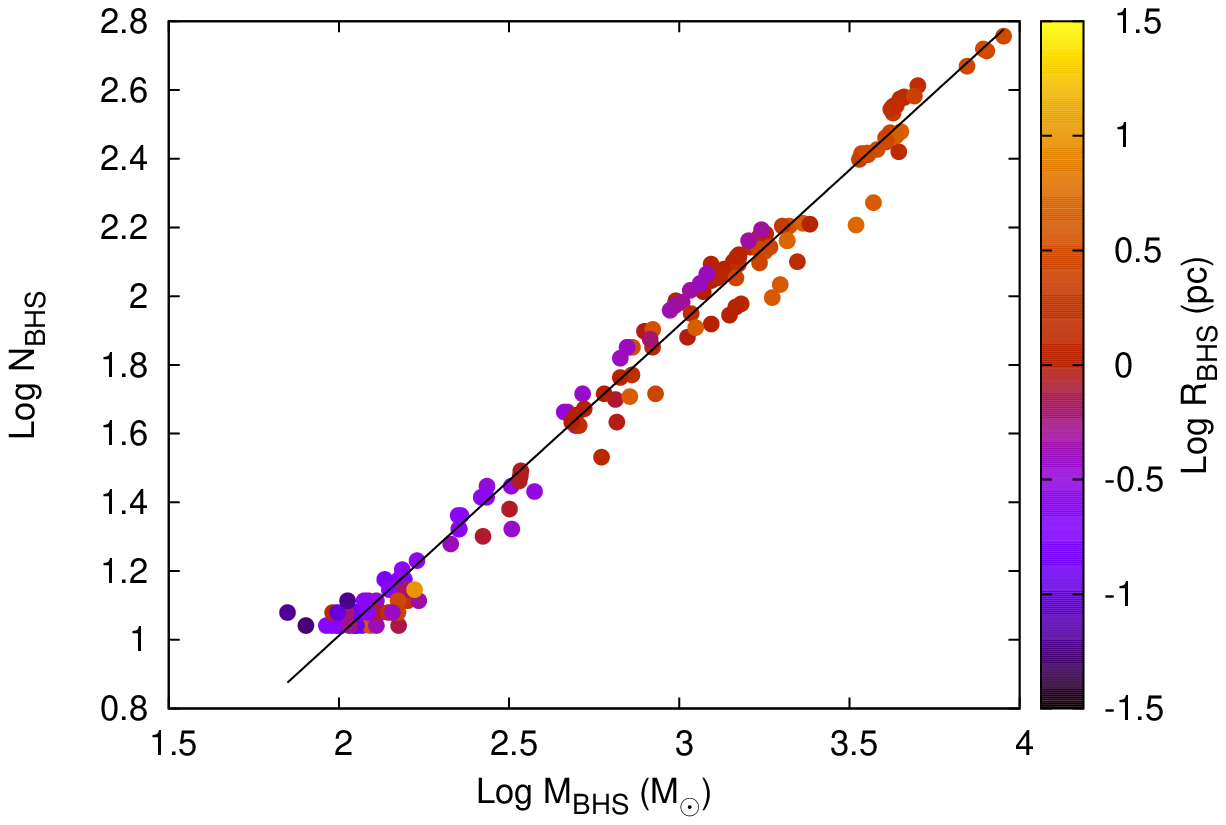}\\
\includegraphics[width=8cm]{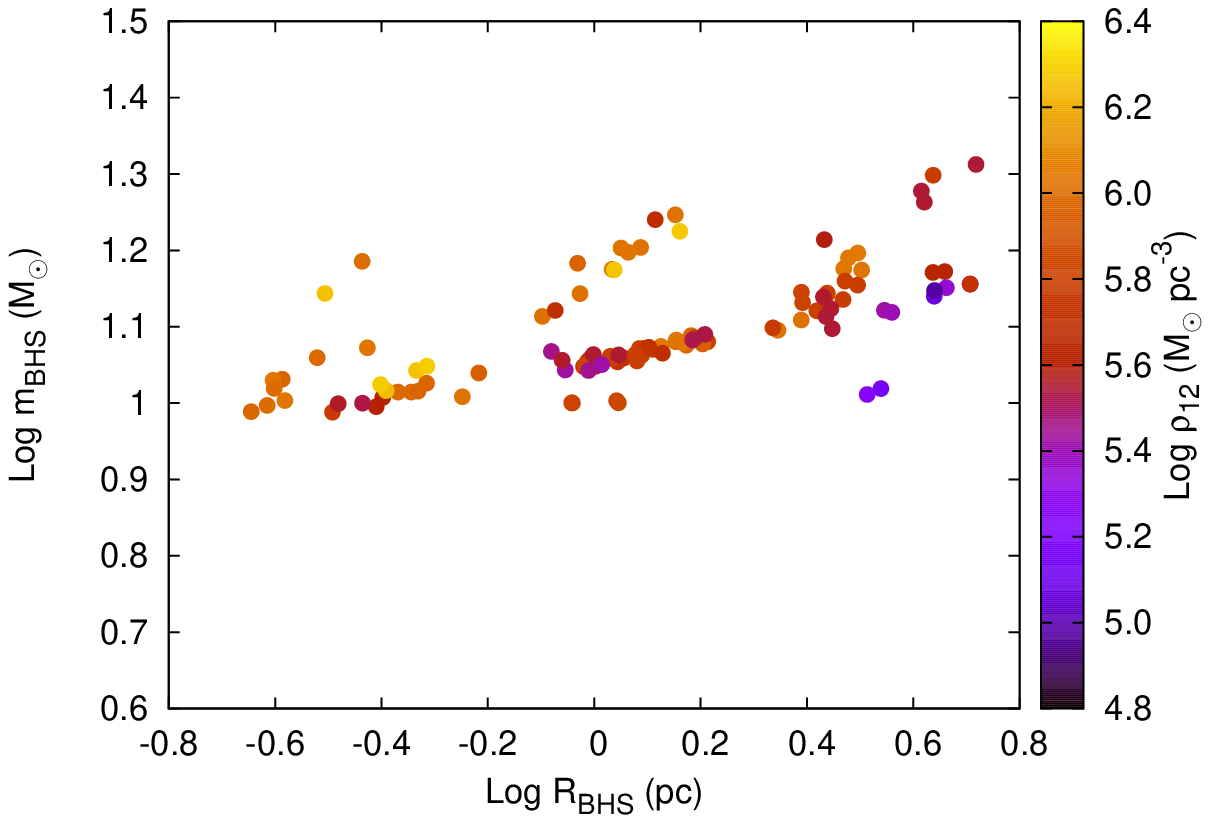}
\includegraphics[width=8cm]{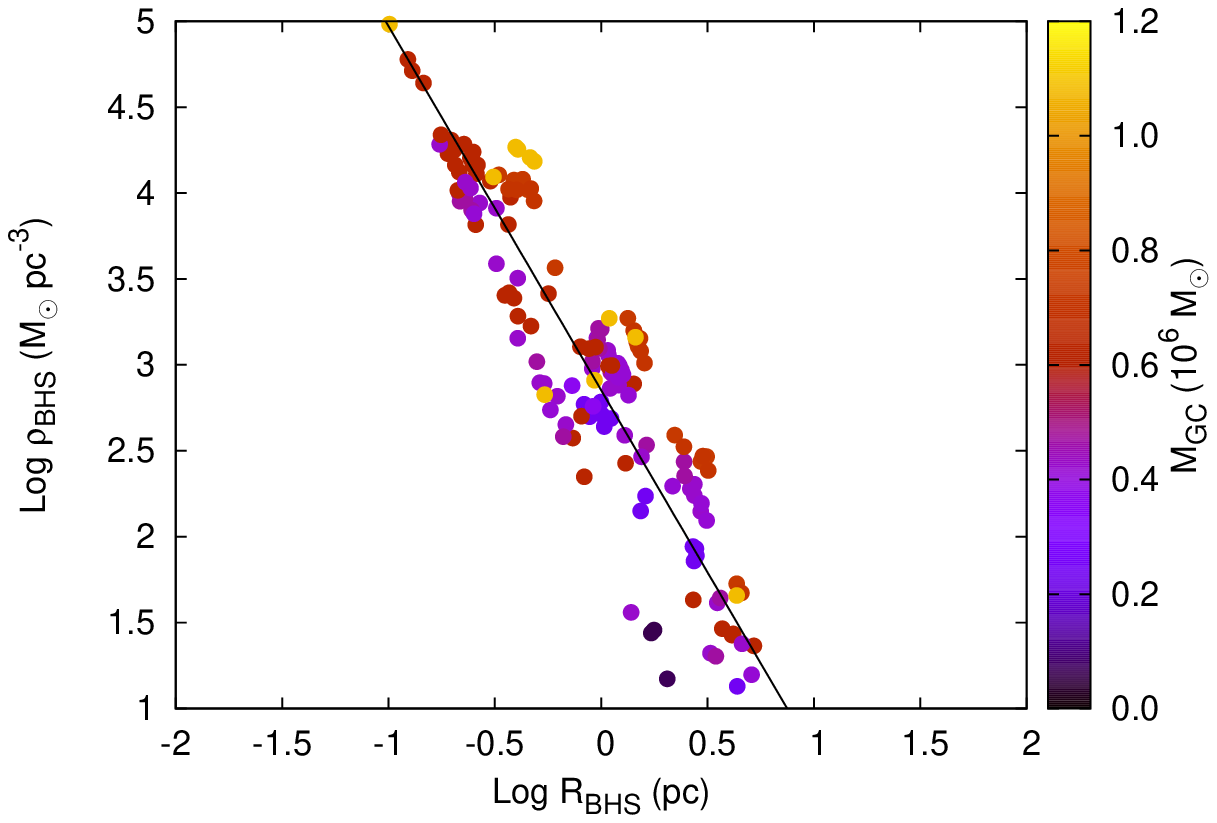}\\
\caption{BHS main correlations. Top left panel: $M_\bhs -R_\bhs$ relation, the color-coded map identifies the host cluster mass at 12 Gyr. Top right panel: number of BHs in the BHS as a function of the BHS mass, the coloured map marks the BHS radius. Bottom left panel: average mass of BHs in the subsystem as a function of $R_\bhs$, the color-coded map refers to the GC central density at 12 Gyr. Bottom right panel: BHS density as a function of its size, mapped on the GC mass at 12 Gyr.
}
\label{F3}
\end{figure*}

Figure \ref{F3} shows the basic correlations linking the BHS fundamental parameters. 
These relations allow us to connect the BHS total mass, radius, and typical density each other. As we show below, the latter quantity can be directly connected with the GC observational properties, making scaling relations the most promising tools to explore the BHS-GC connections.

The $M_\bhs$ and $R_\bhs$ relation is well described by a simple power-law, whose best fitting is given by

\begin{equation}
\Log \left(\frac{M_\bhs}{\Ms}\right) = \alpha \Log \left(\frac{R_\bhs}{{\rm pc}}\right) + \beta,
\label{MRBHS}
\end{equation}
with $\alpha = 0.77\pm0.07$ and $\beta = 3.05\pm0.03$, while the BHS density $\rho_\bhs$ is linked to the BHS size through \begin{equation}
\Log \left(\frac{\rho_\bhs}{\Ms ~{\rm pc}^{-3}}\right) = \alpha \Log \left(\frac{R_\bhs}{{\rm pc}}\right) + \beta,
\label{RhoRBHS}
\end{equation}
with $\alpha = -2.11\pm 0.07$ and $\beta = 2.86\pm0.03$.
 
A closer look at the top left panel of Figure \ref{F3} reveals an interesting connection between the BHS and its host cluster. Indeed, it suggests that more massive clusters harbor heavier BHS at fixed $R_\bhs$ values.

The number of BHs in the BHS correlates very tightly with the BHS mass, as shown in Figure \ref{F3}, through a power-law 
\begin{equation}
\Log N_\bhs = \alpha\Log \left(\frac{M_\bhs}{\Ms}\right)+ \beta
\end{equation}
with slope $\alpha = 0.903\pm0.008$ and intercept $\beta = -0.79\pm0.02$. 
This implies a slow increase of $m_\bhs$ at increasing values of the BHS total mass, being $m_\bhs \propto M_\bhs^{0.1}$. This is in a good agreement with the BHS evolution picture presented at the beginning of this Section. 
Note that the correlation becomes tighter at $N_\bhs > 20$, while below this threshold the data points are much more dispersed. Due to this, in the following we will take into account only subsystems containing at least 20 BHs.  

Top right panel in Figure \ref{F3} makes evident that at a fixed $N_\bhs$ value, larger BHS masses correspond to larger BHS sizes. On the other hand, for fixed BHS mass and $N_\bhs > 20$, a lower number of BHs corresponds to a larger BHS size thus suggesting that the larger the BHS average mass, the larger its size.

The BHS mean mass correlates with $R_\bhs$ 
\begin{equation}
\Log\left(\frac{m_\bhs}{\Ms}\right) = \alpha\Log \left(\frac{R_\bhs}{{\rm pc}}\right)+\beta,
\label{mvsR}
\end{equation}
with $\alpha =  0.13\pm0.01 $, and $\beta =  1.083\pm 0.005$ the best fitting values. 

Moreover, it turns out that the BHS structure depends on the GC central density at 12 Gyr, $\rho_{12}$.  Indeed, our analysis suggests that low-density GCs seem to host BHS characterised by larger $R_\bhs$ values and comprised of heavier BHs than denser GCs, on average.

\subsection{Dynamical consequences of BHS in GCs: phenomenological relations}

\begin{figure}
\centering
\includegraphics[width=8cm]{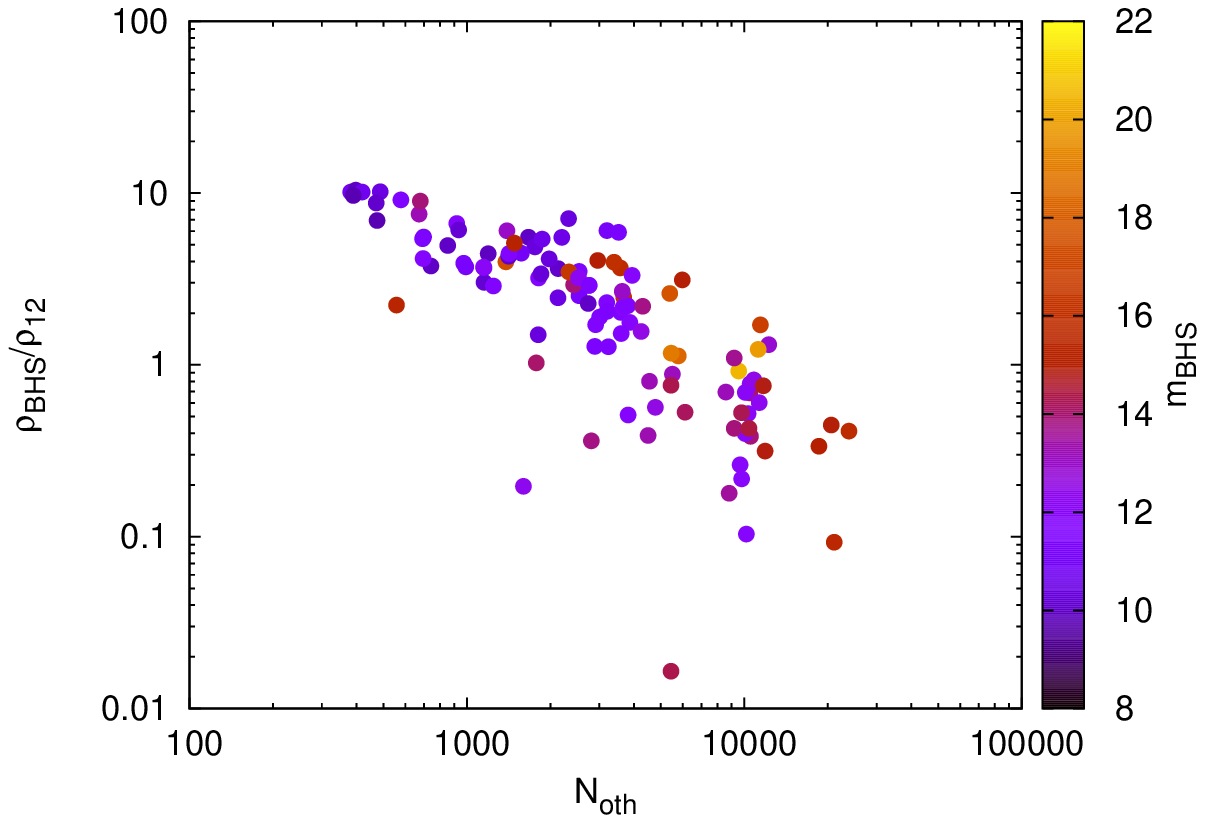}
\includegraphics[width=8cm]{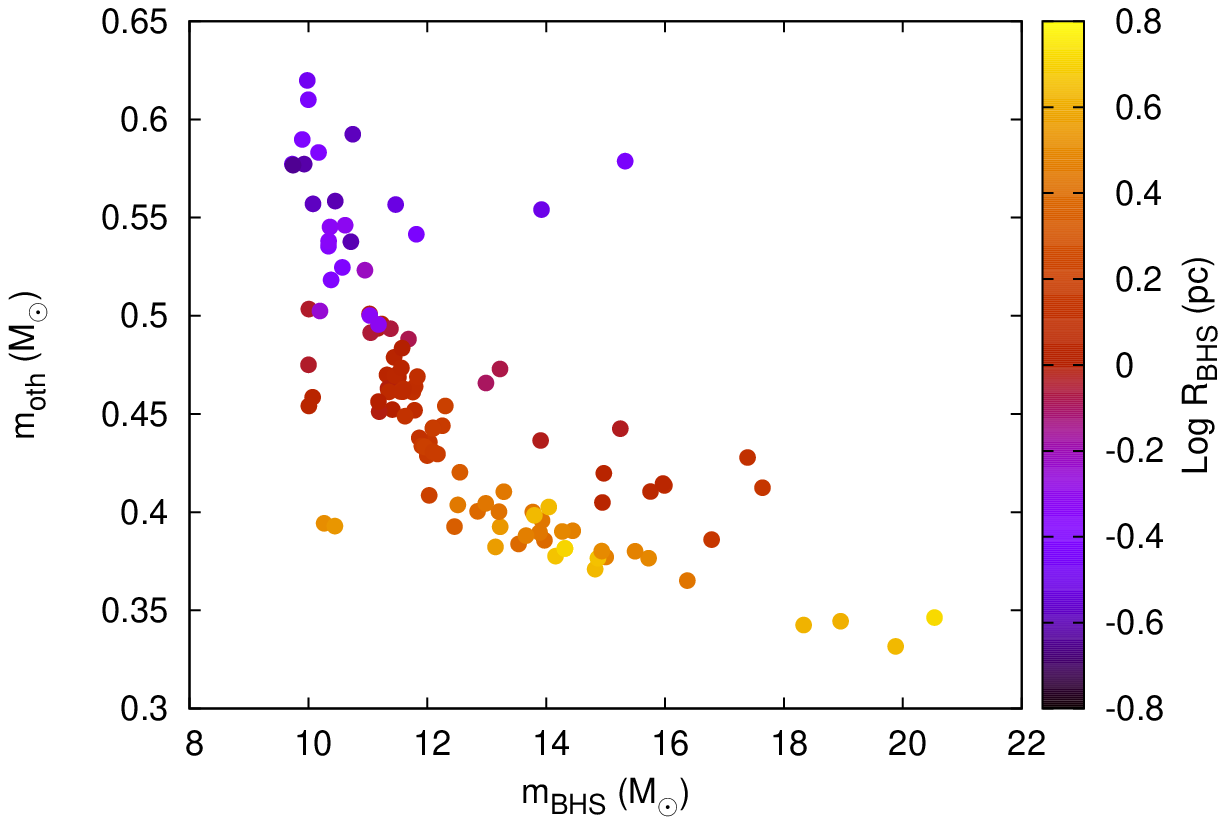}
\includegraphics[width=8cm]{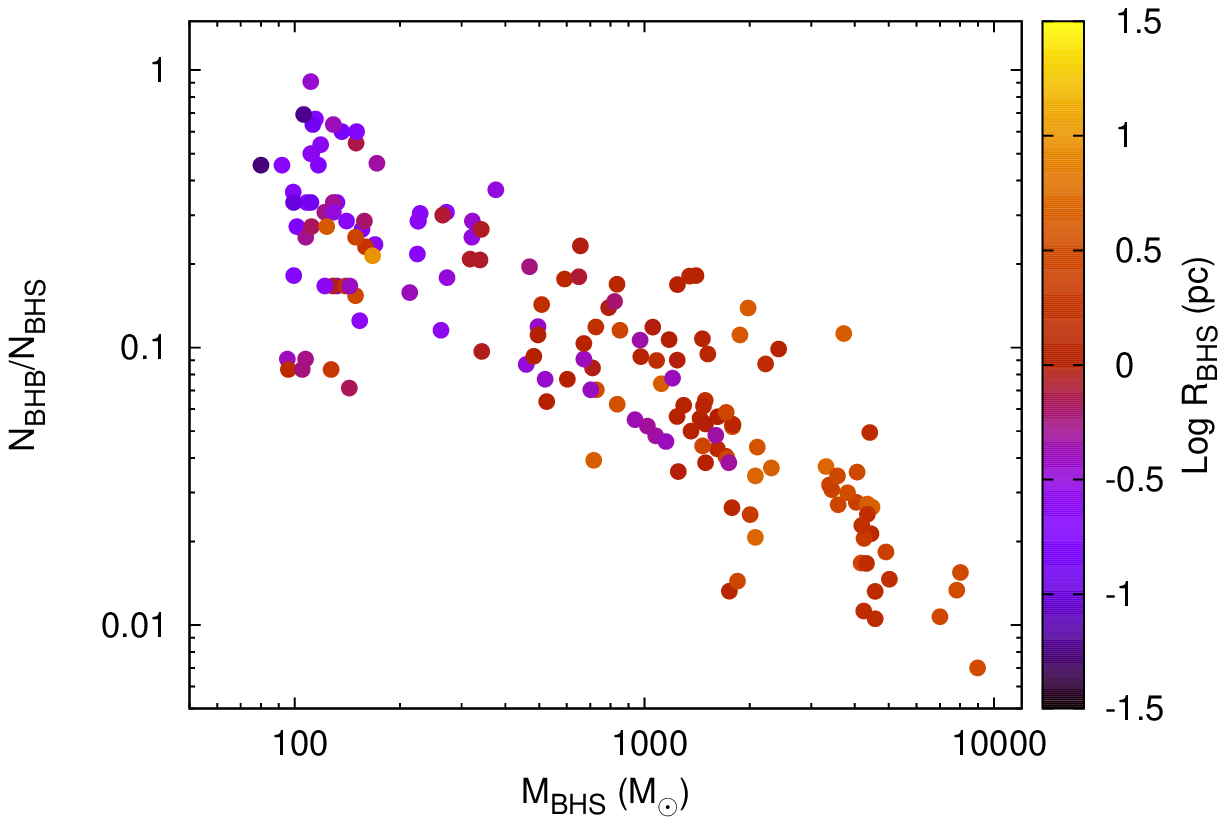}
\caption{Top panel: Central BHS density, normalized to that of the host cluster at 12 Gyr, namely $\rho_{12}$, as a function of the number of stars moving inside $R_\bhs$. The coloured map highlights the BHS average mass.
Central panel: Mean mass of stars contained within $R_\bhs$, as a function of the BHS average mass. The color-coded map represents the BHS size.
Bottom panel: Ratio between the fraction of binaries containing at least one BH and the number of BHs in the BHS, as a function of the BHS mass. The coloured map represents the BHS size.
}
\label{F4}
\end{figure}

In this section, we will investigate whether our BHS definition can be used to connect the GC dynamical status with the retained BH population. Indeed, the presence of a conspicuous number of BHs surviving into the host cluster core up to 12 Gyr is expected to shape significantly the GC properties. 
\\
For instance, the top panel of Figure \ref{F4} shows how the BHS and GC densities vary at varying the number of stars within $R_\bhs$ and the BHS average mass. 

This relation can hide some information about the dynamical status of the host GC which is rather difficult to see.

We can schematize the GC-BHs common evolution and ``dynamical feedback'' as follows:
\begin{enumerate}
\item massive stars evolve and become BHs while rapidly segregating to the GC core, leading to the formation of a massive BHS;
\item the BHS injects energy in the surroundings, losing energy to other stars and causing the GC core expansion, thus leading to a lower GC central density;
\item the formation of massive BH-BH binaries in the BHS provides a sufficient energy supply to sustain the GC core, leading eventually to its expansion;
\item repeated strong single and binary encounters occurring inside the BHS drive the ejection of the most massive BHs and stars, causing the BHS contraction due to the loss of the energy supply. Consequently, the mean BHS mass and size decrease while its density increases. 
\end{enumerate}

The scaling relation presented here seem to be compatible with the above scheme, allowing us distinguishing between ``dynamically young'' massive and relatively loose BHS that inhabit dense GCs, and ``dynamically old'' BHS, lighter, denser and inhabiting GCs characterized by smaller central densities.

Hence, it appears evident a correlation between the BHS-GC density ratio and the number of stars mixed with BHs inside the BHS radius. This implies that there is a relation between the potentially observable stellar properties and the BHs composing the BHS.

Indeed, the BHS average mass is tightly connected with the average mass ($m_\oth$) of stars enclosed within $R_\bhs$, as shown in the central panel of Figure \ref{F4}, through the relation
\begin{equation}
\left ( \frac{m_\oth}{\Ms} \right) = \alpha \frac{1 + \beta\left (m_\bhs/\Ms \right)}{1 + \gamma\left(m_\bhs/\Ms \right)},
\label{Eq1}
\end{equation}
with $\alpha =  0.2 \pm 0.1$, and $\beta = -0.19\pm0.09$ and $\gamma = -0.14\pm0.02$.

More interestingly, the central panel in Figure \ref{F4} illustrates that at increasing $R_\bhs$ values, BHS host heavier BHs and lighter stars. 
This implies that the larger the BHS average mass, the larger the number of stars ``mixed'' with the BHs in the subsystem, since $N_\oth m_\oth = N_\bhs m_\bhs$ by definition.

 The relations found between ordinary stars and BHs in the BHS, together with the relation between the GC and the BHS central density, suggest that BHS hosting heavy stellar BHs have, on average, a low density concentration.

In dynamically young GCs, the BHS is large and sparse and its BHs have large masses. In these ``active systems'', BHs did not have enough time to contract sufficiently and form a dense BHS, while their self-interactions, which are the main engine for the ejection of massive BHs, did not become effective yet. Consequently, a large population of heavy BHs move inside the GC after 12 Gyr of evolution.
Hence, to provide a sufficient energy flow at the half-mass radius, only a small number of BH binaries is needed, being these extremely efficient energy sources that can lead to large GC half-mass radii. 

At a fixed value of the semi-major axis, the heavier the binary the larger the binding energy. Under the general assumption that binaries binding energy undergoes a nearly constant variation ($\Delta(Eb)/Eb = -0.4$, \cite{heggie75} $\Delta(Eb)/Eb = -0.2$, \cite{spitzer87}), the heaviest binary BHs represent the most effective energy source in the cluster.
However, decreasing the binary mass from $m_{b1}$ to $m_{b2}$ implies that 
the number of interactions needed to produce the same amount of energy must increase by a factor ($m_{b1}/m_{b2})^{2}$, which in turn implies larger densities. Ejection of the most massive BH binaries due to dynamical interactions leads to the contraction of the BHS in order to increase the energy generation by lower mass BH binaries. The more compact and dense the BHS is, the higher the number of interactions that are needed to sustain the energy flow through the half-mass radius. As a consequence, for dynamically older systems, BHS are denser, more compact and with smaller mass BHs.
Figure \ref{F4} demonstrates that GCs with more massive BHS have fewer number of their BHs in binary systems. Moreover, decreasing values of the $N_{\rm BHB}/N_\bhs$ ratio correspond to an increase of the BHS size. This is supported by the bottom panel of Fig. \ref{F4}, which shows how the ratio between the number of BHs in binary system and those in the BHS varies at varying the BHS mass and its radius. Indeed, heavier and larger BHS are characterised by a lower fraction of binary systems.

 Our results compares very well with \cite{breen13} predictions. Clusters with larger half-mass relaxation times can sustain long living and more massive BHS than clusters with smaller half-mass relaxation times, for which the BHS contracts much faster and ``burns'' the more massive BHs that are needed to generate the required energy to support the host cluster.

\begin{figure}
\centering
\includegraphics[width=8cm]{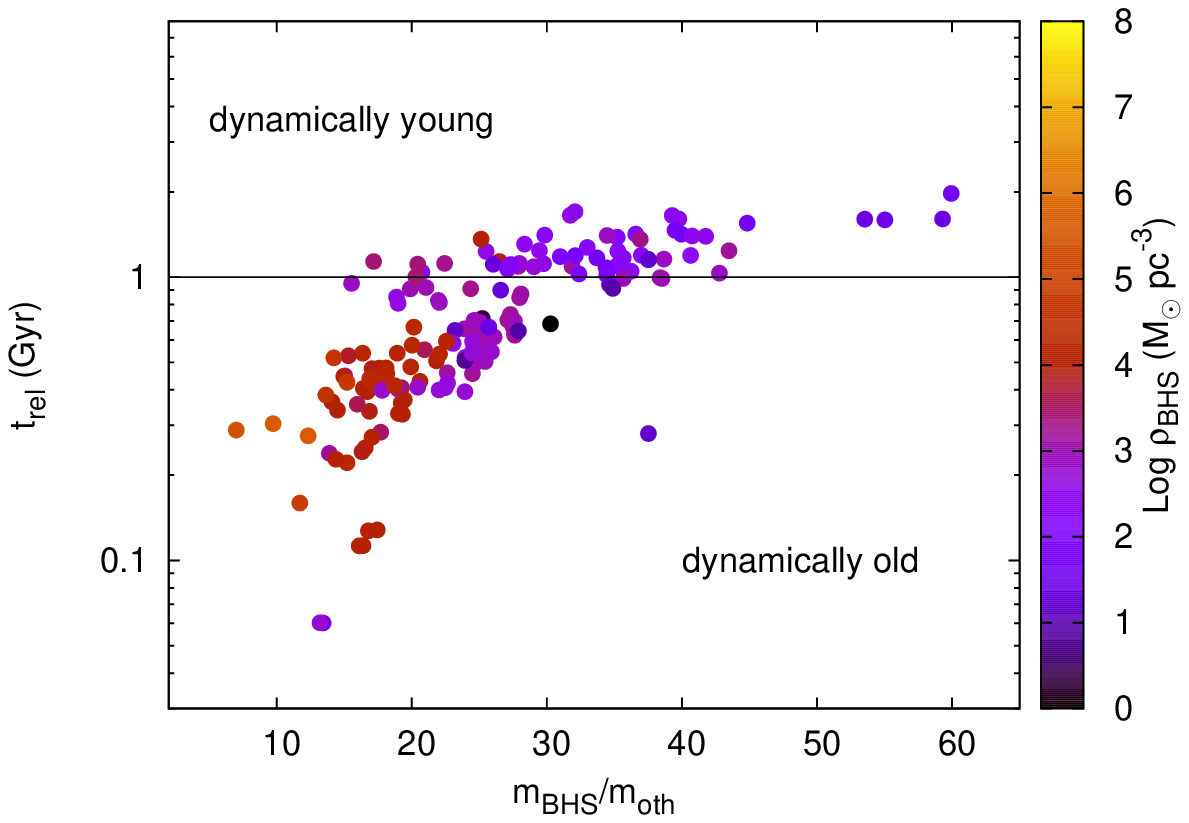}\\
\includegraphics[width=8cm]{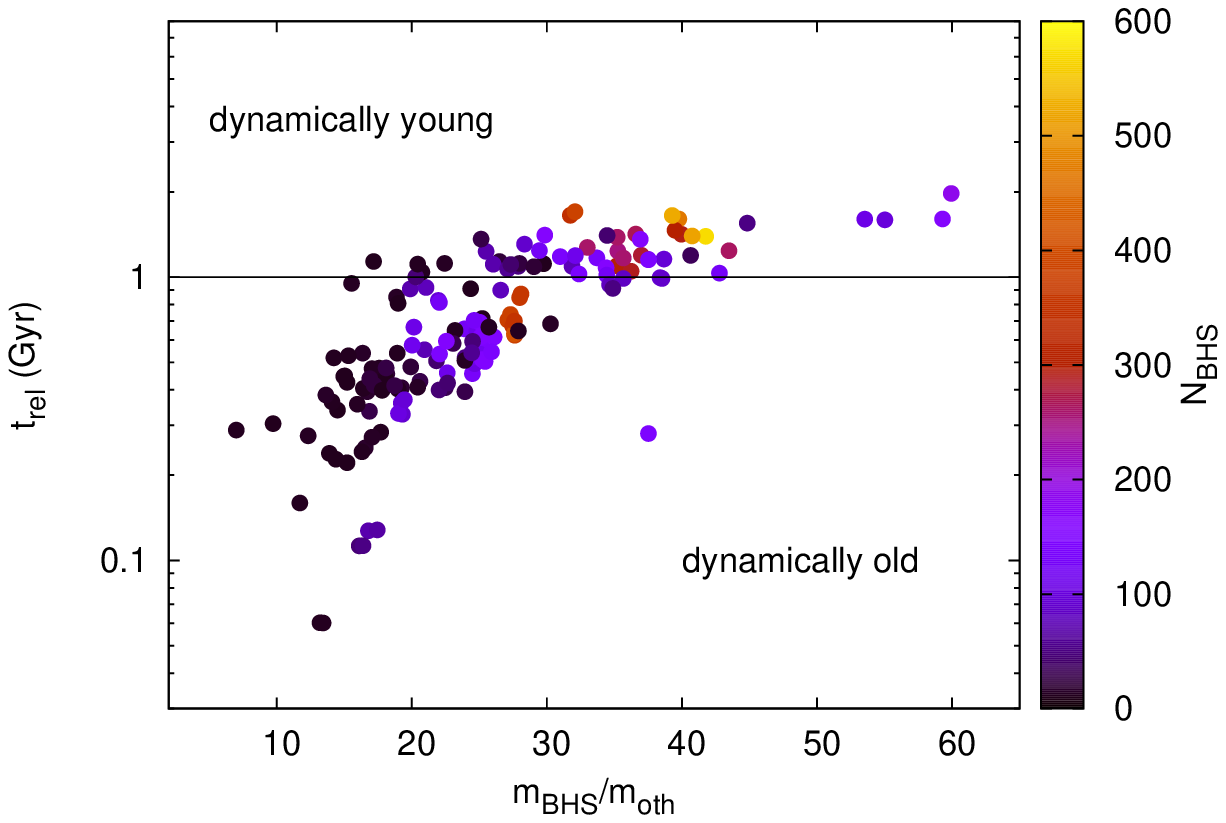}\\
\includegraphics[width=8cm]{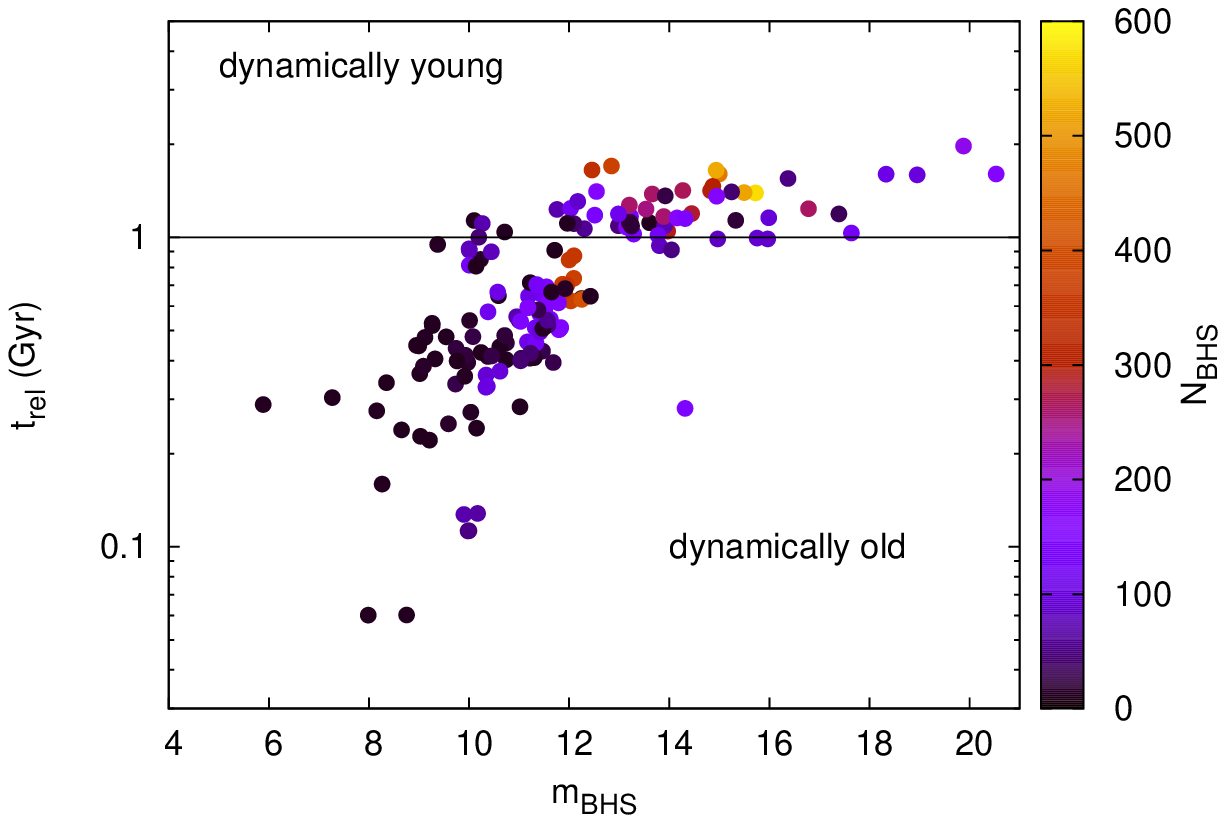}
\caption{Top panel: Host cluster relaxation time-scale at 12 Gyr as a function of the ratio between stars and BHs averaged mass, calculated inside $R_\bhs$. The coloured map labels the BHS central density. Central panel: the same as in the top panel, but here the color-coded map represents the number of BHs in the subsystem. Bottom panel: the same as in the top panel, but on the X-axis is shown the BHS mean mass.}
\label{F7}
\end{figure}

 Figure \ref{F7} shows how the GC relaxation time at 12 Gyr ($t_{\rm rel}$) varies with the ratio between the BHS and mixed stars average masses. Here, we used the standard definition of half-mass relaxation time-scale defined so far by \cite{spitzer87}. 
The three panels in Figure \ref{F7} outlines that the densest BHS, characterised by a smaller ratio between $m_\bhs$ and $m_{\rm oth}$ on average, are found in GCs characterized by low $t_{\rm rel}$, thus dynamically old at 12 Gyr. On another hand, dynamically younger systems host low-density BHS, containing a significant fraction of massive stellar BHs and in general, a larger number of BHs, as shown in the bottom panel of Figure \ref{F7}.

\begin{figure}
\centering
\includegraphics[width=8cm]{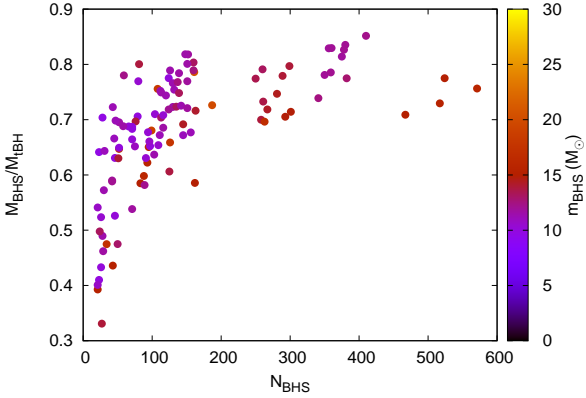}\\
\caption{BHS mass, normalized to the mass of the whole population of BHs in the cluster as a function of $N_\bhs$. The color-coded map highlights the BHS average mass. }
\label{F8}
\end{figure}

Surprisingly, we found that our definition of BHS has crucial implication on the relation between the BHS and the whole BH population in a GC. Figure \ref{F8} shows the ratio between the BHS mass and the total mass of retained BHs after 12 Gyr as a function of the number of BHs in the subsystem. As long as the number of BHs in the subsystem remains below $\sim 100$, we found that the BHS contains up to $70\%$ of the whole BHs population. For subsystems containing a larger number of BHs, instead, this percentage oscillates betwee $70-85\%$.

The $M_\bhs-M_{\rm tBH}$ is a simple power-law
\begin{equation}
\Log \left(\frac{M_{\rm BHS}}{\Ms}\right) = \alpha \Log \left(\frac{M_{\rm tBH}}{\Ms}\right)+ \beta,
\end{equation}
where in this case $\alpha = 1.14\pm 0.02$ and $\beta = -0.62 \pm 0.06$.
Hence, our procedure allows calculating the mass of a central BH subsystem from the knowledge of the whole population of BHs present in the cluster at that time. 

\subsection{BHS observational scaling relations}
\label{sec:obser}

A challenging quest is to determine the presence of a BHS in the interior parts of GCs. For this purpose, we extracted from our MOCCA models the GC central velocity dispersion, $\sigma$,  total luminosity $L$, observational half-mass radius $r_{\rm h,obs}$, and central surface brightness $\Sigma$. 
Our aim is to provide a set of scaling relations that can be used to infer the presence of a BHS in any given GC for which these global observational properties are known.

In the following expressions, we will infer the BHS-GC observational correlations with the general expression (if not specified otherwise),
\begin{equation}
\Log \rho_\bhs = A \Log {\rm X} + B,
\end{equation}
where $X$ is the observational parameter considered. We use letters $A$ and $B$ instead of $\alpha$ and $\beta$ to better highlight the difference between observational and structural, or ``dynamical'', correlations.
Also, we will only consider models having at least 20 BHs within $R_\bhs$ as done in the previous sections (unless specified differently).

\begin{figure}
\centering 
\includegraphics[width=8cm]{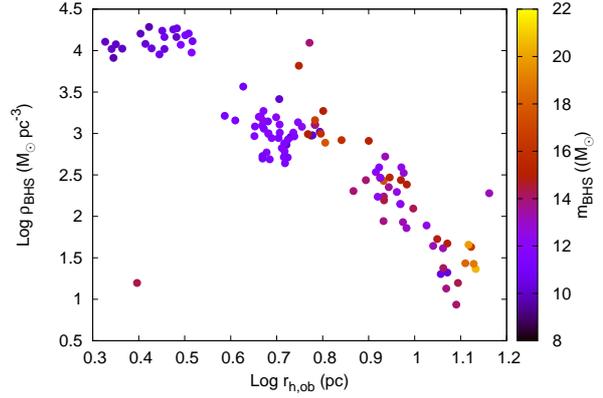}\\
\caption{BHS density as a function of the GC observational half-mass radius. The color coded map represents the BHS average mass.}
\label{radii}
\end{figure}

The BHS mean density seems to weakly correlate with the $r_{\rm h,obs}$, as shown in Figure \ref{radii} through a power-law 
\begin{equation}
\Log \left(\frac{\rho_\bhs}{\Ms {\rm pc}^{-3}}\right) = A \Log \left(\frac{r_{\rm h,obs}}{{\rm pc}}\right) + B,
\end{equation}
with intercept $B = 5.5\pm 0.1$ and $A = -3.4\pm 0.2$.

\begin{figure}
\centering 
\includegraphics[width=8cm]{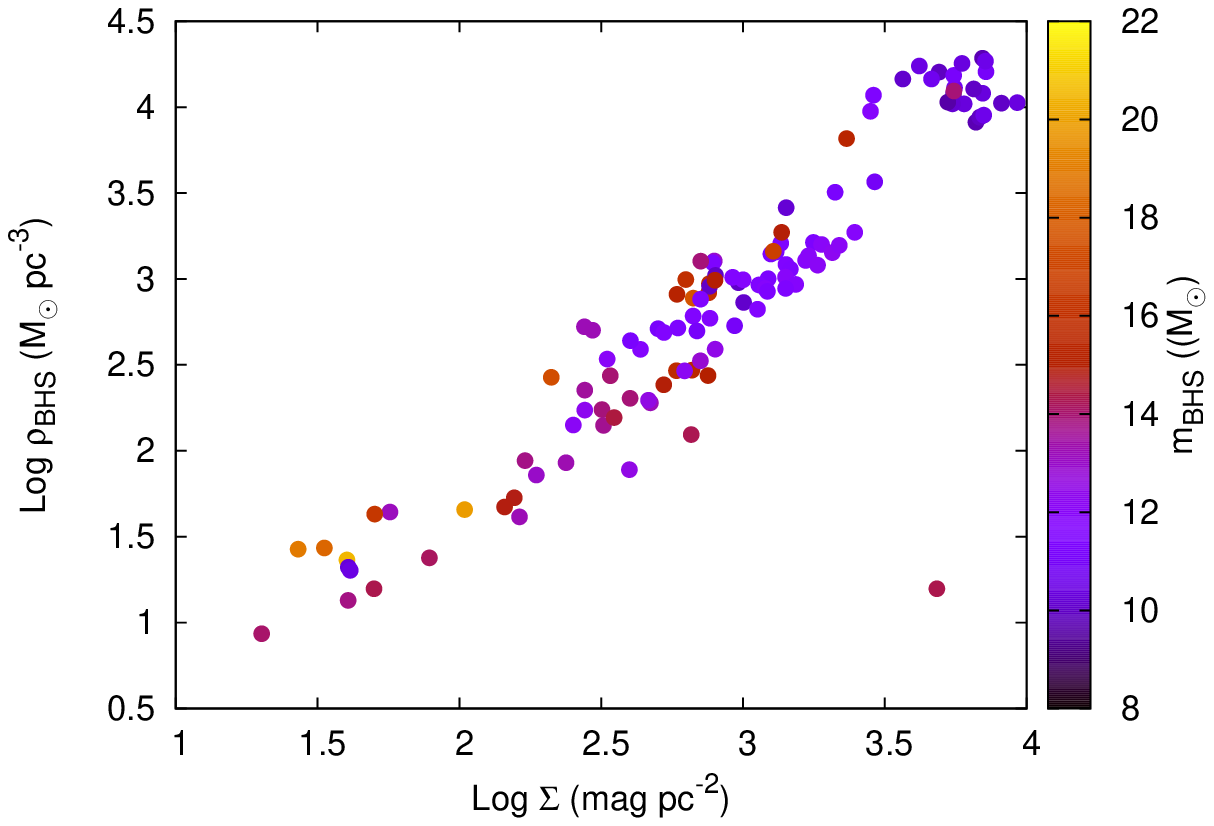}\\
\includegraphics[width=8cm]{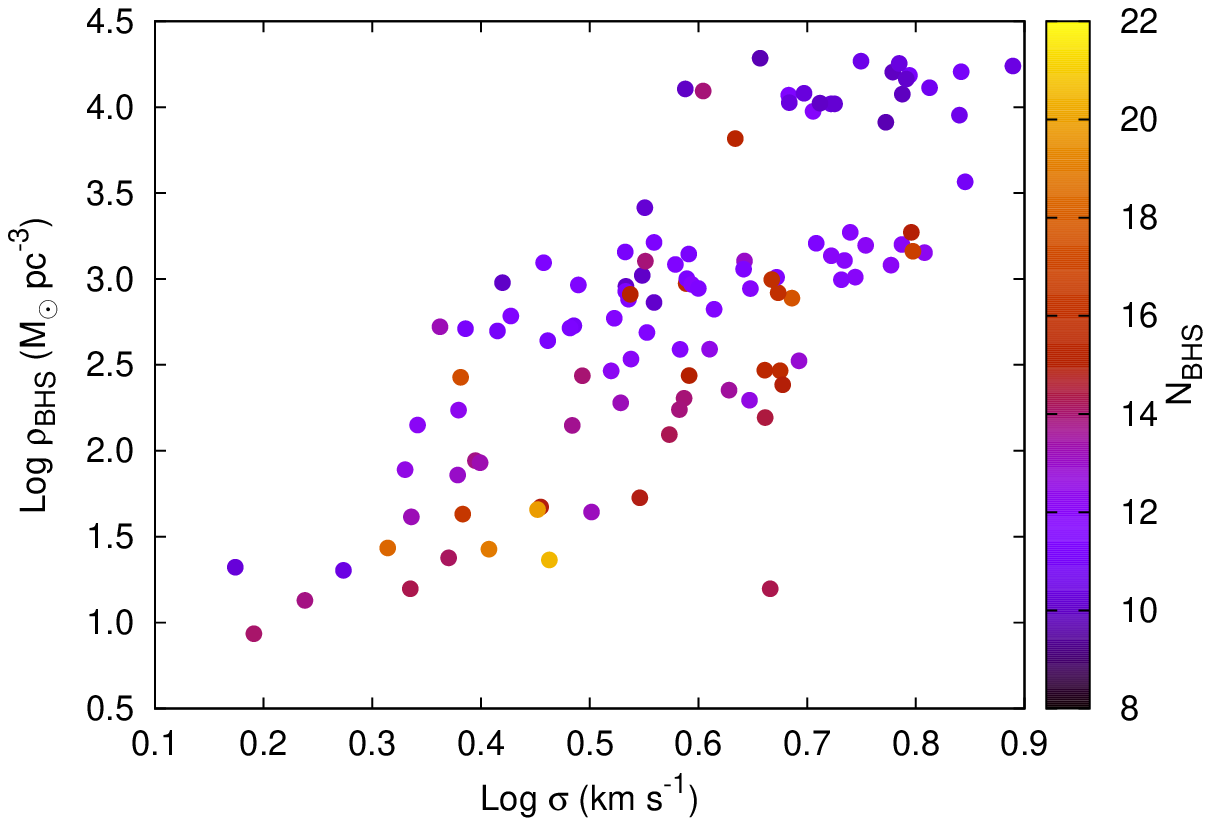}\\
\includegraphics[width=8cm]{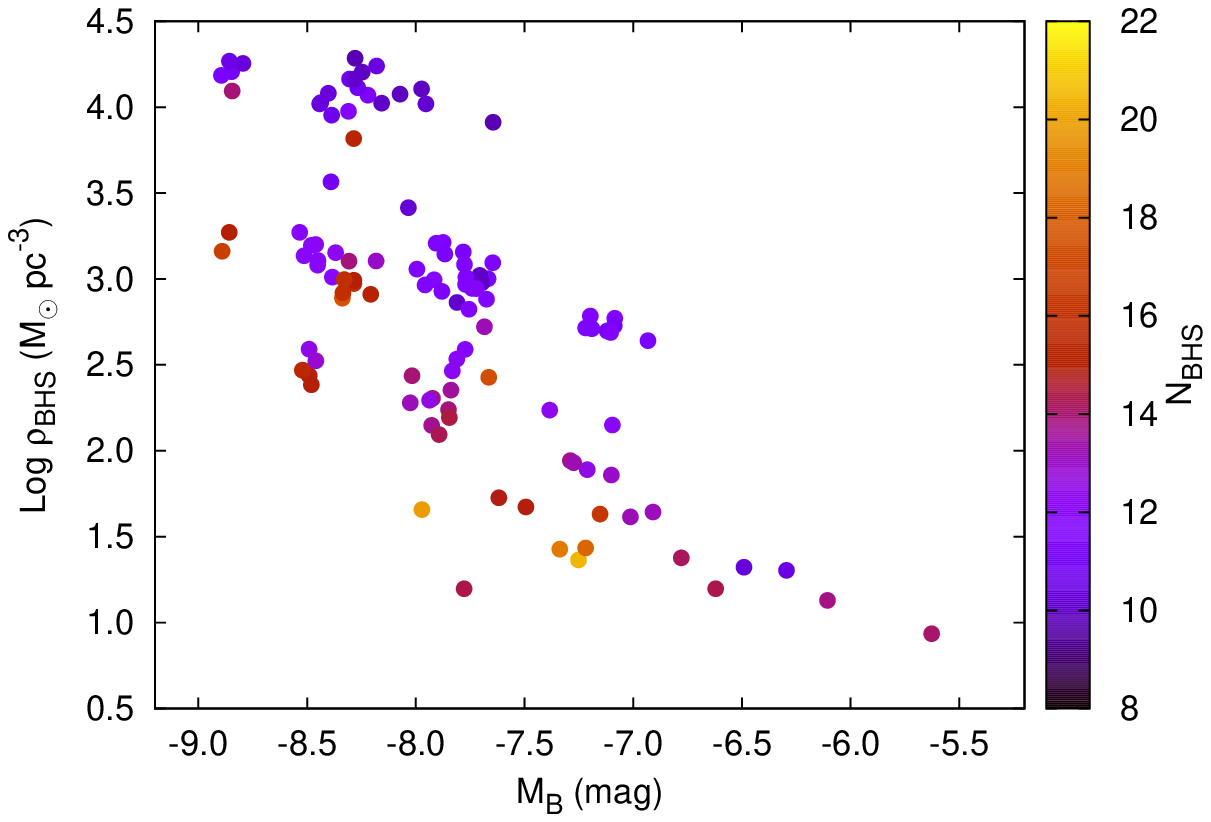}
\caption{BHS scale density as a function as a function of the GC central surface brightness (top panel), central velocity dispersion (central panel) and magnitude (bottom panel). The color-coded map identifies the number of BHs in the BHS.}
\label{glob}
\end{figure}

In general, the correlation between the BHS density and the global GC properties are not very tight. From Figure \ref{glob}, it appears evident that neither the total magnitude in the B-band or the GC velocity dispersions are good indicators for extracting information about the BHS density. 

In an attempt to define a set of correlations capable to link the BHS properties and several GC observables, we find a ``fundamental plane'' for BH subsystems that allows us to connect the BHS density with the GC average surface luminosity, namely $L/r_{\rm h,obs}^2$, and its velocity dispersion, $\sigma$. 

\begin{figure}
\centering
\includegraphics[width=8cm]{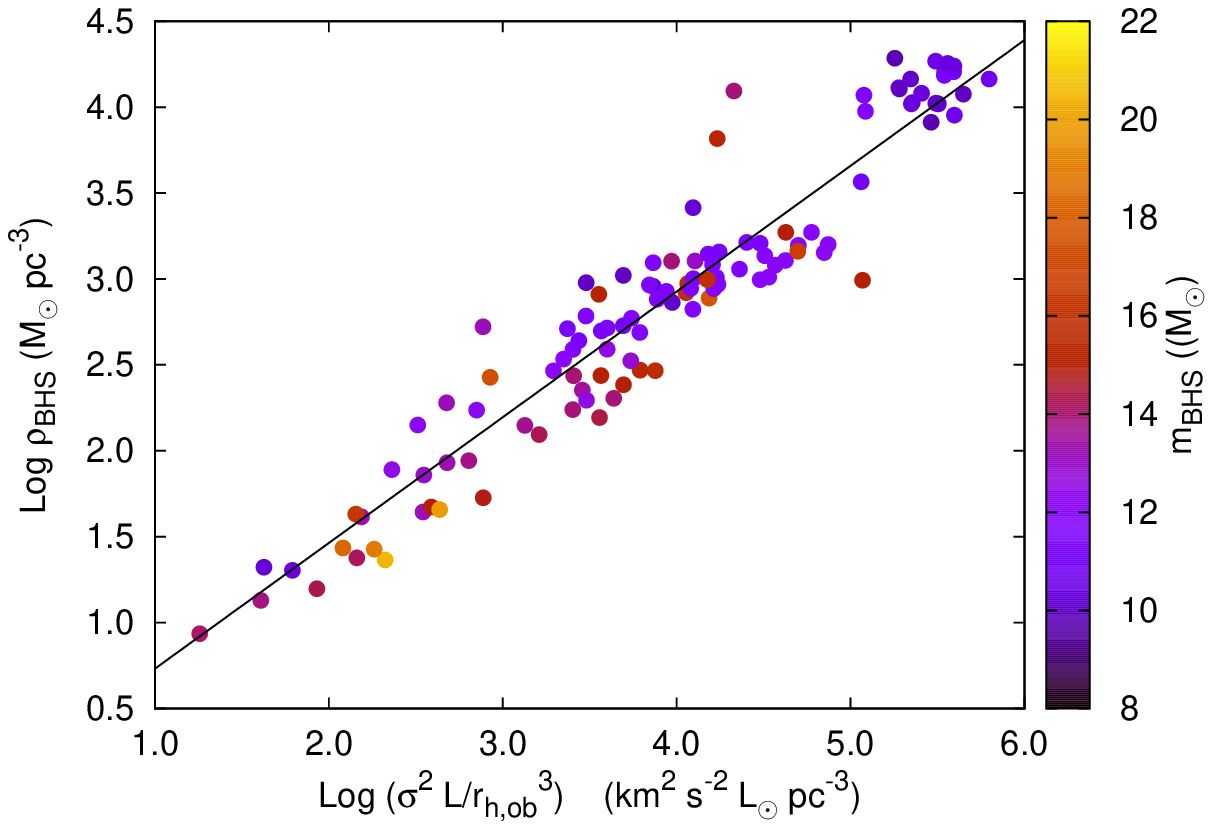}
\includegraphics[width=8cm]{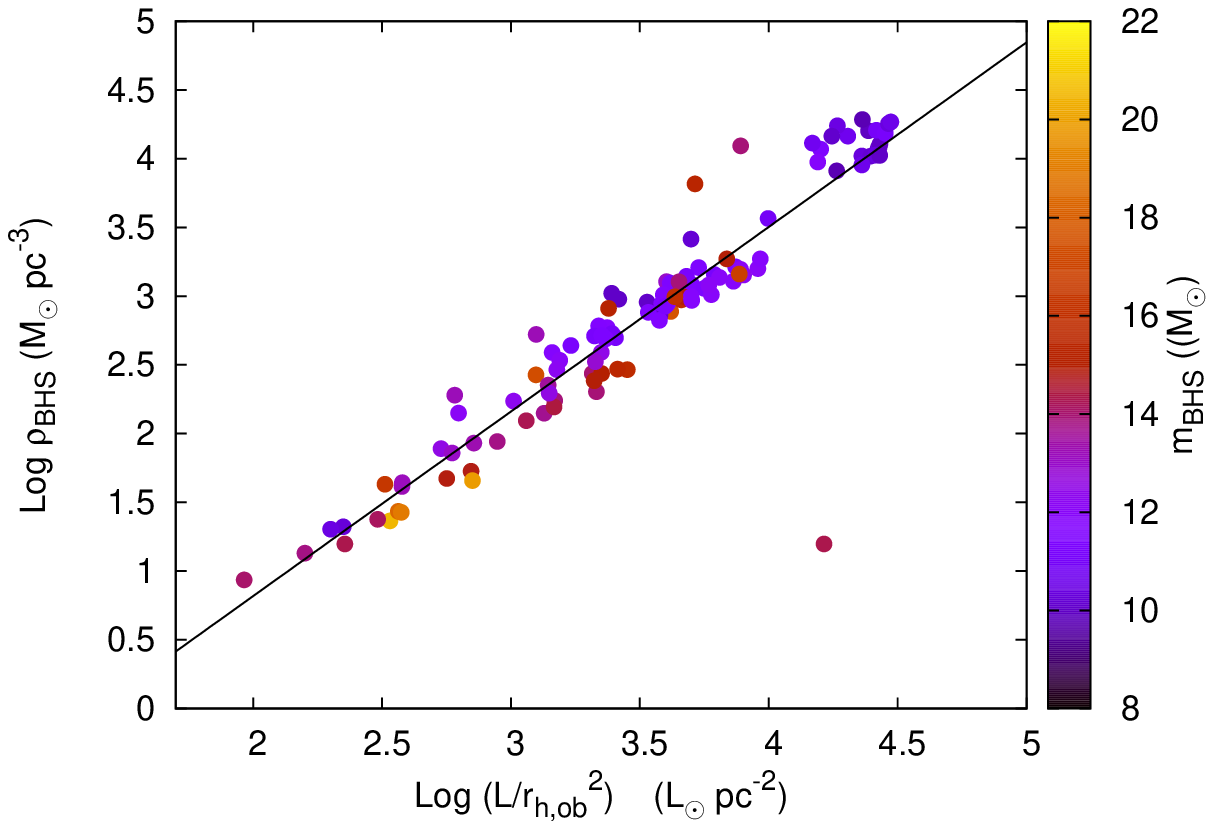}
\caption{Fundamental plane (top panel) and reduced fundamental plane (bottom panel) 
for BHSs. The color-coded map refers to the BHs mean mass in the BHS.}
\label{fund}
\end{figure}

This correlation, shown in the top panel of Figure \ref{fund}, suggests that the lower the GC average luminosity density (the term $L/r_{\rm h,obs}^3$) and its mean kinetic energy ($\sigma^2$) the lower the BHS density. Note that on average, low-density BHS have larger mean masses. 

A much tighter relation, as shown in the bottom panel of Figure \ref{fund}, can be used simply combining $\rho_\bhs$ with the GC average surface luminosity $L/r_{\rm h,obs}^2$. 
In this case, the relation can be written as
\begin{equation}
\Log \left(\frac{\rho_\bhs}{\Ms {\rm pc}^{-3}}\right) = A\left[\Log \left(\frac{L}{{\rm L}_\odot}\right)-2\Log \left(\frac{r_{\rm h,obs}}{{\rm pc}}\right)\right]+B,
\label{fun}
\end{equation}
with $A = 1.34\pm 0.05$ and $B = -1.87\pm0.18$. This very simple relation implies that the total GC luminosity and its observational half-mass radius can be used to obtain the BHS density. Once $\rho_\bhs$ is obtained, the BHS size and mass can be obtained using 
Equation \ref{RhoRBHS} and \ref{MRBHS}, respectively.

\section{A fundamental plane for IMBHs}

The procedure described above allowed us to define a handful of scaling relations connecting the GCs observational properties and their BHs population. In order to determine whether our treatment can be used also for IMBHs, we grouped all the MOCCA models harboring a central BH with a mass above $150\Ms$ at 12 Gyr. 

In this case, to define the typical ``IMBH size'', we made use of the widely known concept of influence radius, $R_\ibh$, which is the region where the IMBH dominate the dynamics \citep{merritt04b}.

\begin{figure}
\centering
\includegraphics[width=8cm]{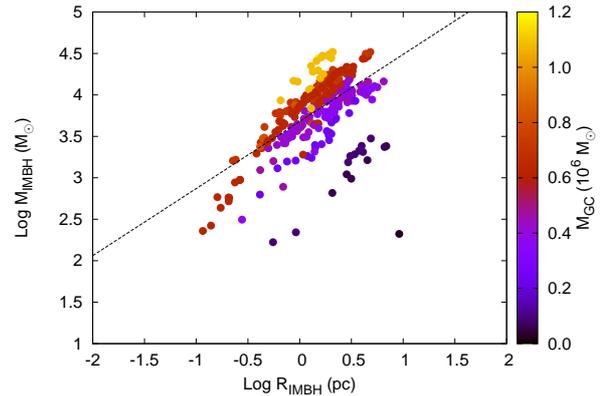}
\caption{IMBH mass as a function of the influence radius. The color-coded map identifies the host GC final mass. }
\label{RMibh}
\end{figure}

Similarly to BHS, the IMBH mass, $M_\ibh$, is connected to $R_\ibh$ through a power-law
\begin{equation}
\Log M_\ibh = \alpha\Log R_\ibh + \beta,
\label{Eibh1}
\end{equation}
with $\alpha = 0.81\pm0.06$ and $\beta = 3.68\pm0.02$, close to the values obtained for BHS. The corresponding relation is shown in Figure \ref{RMibh}.

The similarity between Equation \ref{Eibh1} and \ref{MRBHS} suggests that BHS acts like a central point-like mass, shaping significantly the mass distribution in the inner regions of the parent cluster.

The IMBH mass and radius can be combined to define a typical density
\begin{equation}
\rho_\ibh = M_\ibh / R_\ibh^3,
\end{equation}
which can be used to connect the GC ``dark'' properties and its observational parameters.

Even for IMBHs, is possible to define a fundamental plane, defined by $\rho_\ibh$ and the GC typical surface luminosity, defined as the ratio between the total bolometric luminosity and the square of the half-light radius. Analogously to Eq. \ref{fun}, the fundamental plane is well described by a power-law. 
\begin{equation}
\Log \left(\frac{\rho_\ibh}{\Ms {\rm pc}^{-3}}\right) = A\left[\Log \left(\frac{L}{{\rm L}_\odot}\right)-2\Log \left(\frac{r_{\rm h,obs}}{{\rm pc}}\right)\right]+B,
\label{fun}
\end{equation}
with 
\begin{align}
A &= 1.35\pm0.04\\
B &= -2.3\pm0.2 ~.
\end{align}

Therefore, it seems that a strategy similar to the one applied to BHS could successfully be used to infer the basic properties of IMBHs and their environments on the basis of the host cluster observational parameters.

A deeper and more detailed analysis of how using our model to target GCs potentially harbouring an IMBH will be provided and discussed in our forthcoming paper.

\begin{figure}
\centering
\includegraphics[width=8cm]{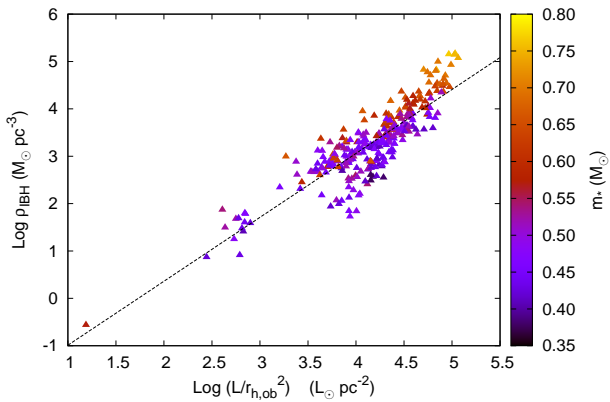}
\caption{
Fundamental plane for IMBHs heavier than $8\times 10^3\Ms$ (filled triangles) and lighter than this limiting value (filled circles). The color-coded map identifies the average stellar mass enclosed within $R_\ibh$.
}
\label{ibhfun}
\end{figure}

\section{Discerning BHSs, IMBHs and ordinary stars in GC centre}
\label{sec:disce}

The results discussed above allow us to obtain a simple set of scaling relations connecting GCs observable quantities and the main structural parameters characterising the population of BHs confined deeply into the host cluster.

In particular, we have shown that the cluster total luminosity and observed half-light radius represent the parameters that mostly constrain the BHS typical density, which in turn can be used to calculate the BHS mass, radius and average mass.
In this section, we investigate whether the quantity $L/r_{\rm h,ob}^2$ and the presence of a BHS can be connected uniquely. Moreover, we try to understand whether is possible to place any constrain on the putative presence of a BHS in Milky Way GCs.

\begin{figure*}
\centering
\includegraphics[width=12cm]{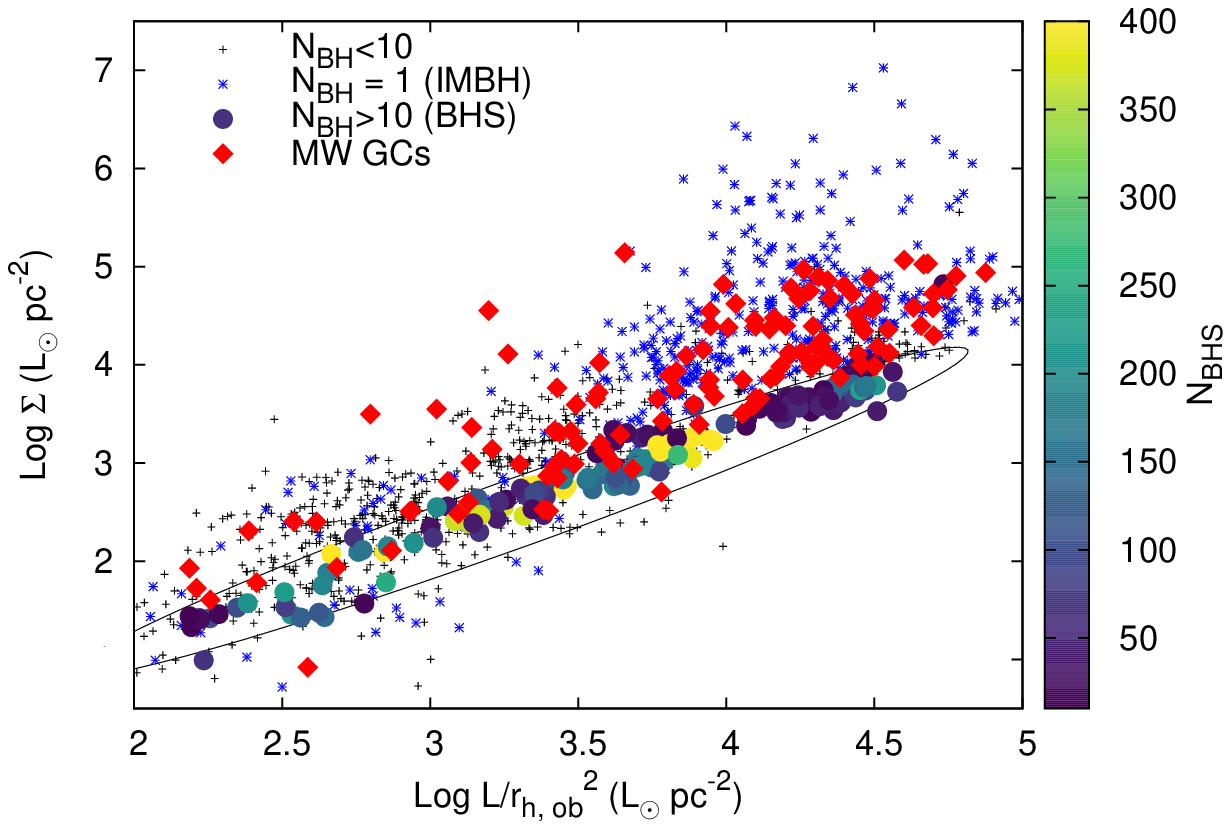}\\
\includegraphics[width=12cm]{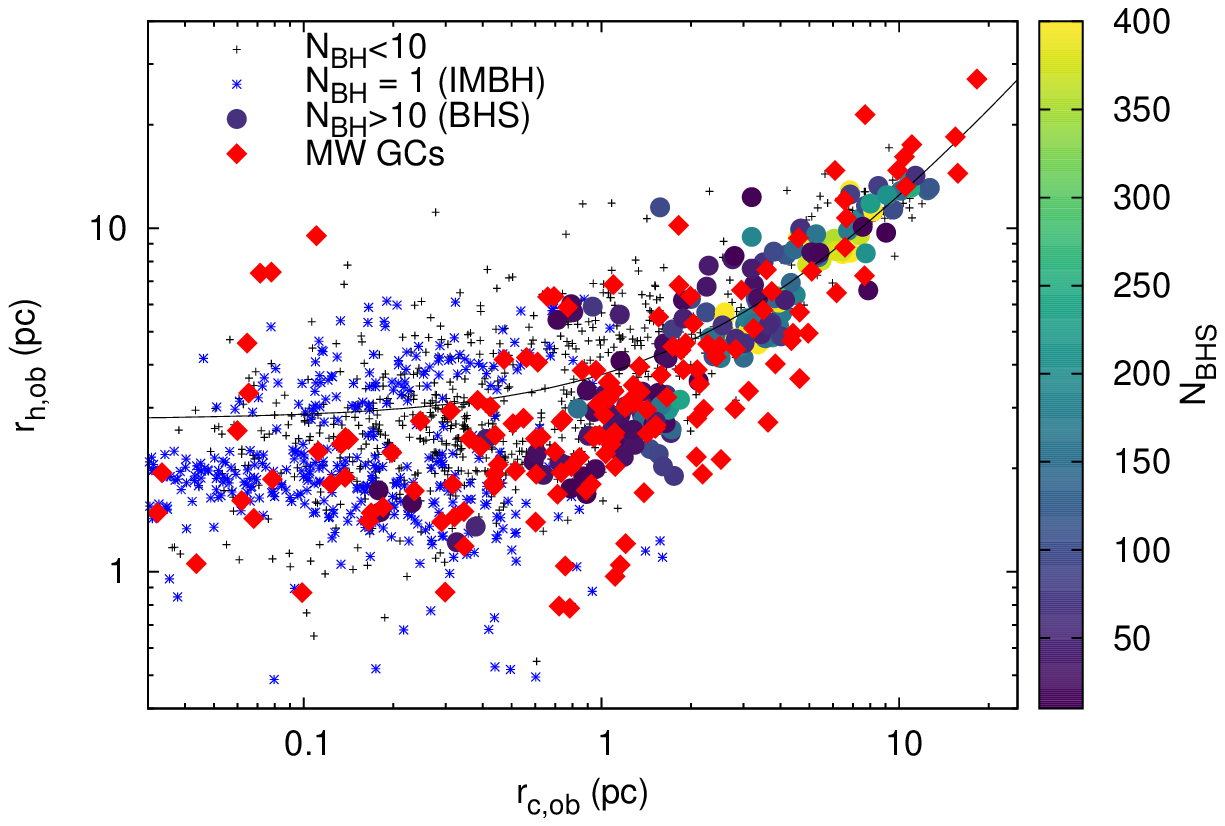}
\caption{Top panel: Central surface brightness as a function of the average surface luminosity, for all the MOCCA models at 12 Gyr and for MW GCs.  Bottom panel: as above, but here we show the observational half-light radius as a function of the observational core radius. In both panels, we distinguish between GCs hosting an IMBH with mass above $150\Ms$ (blue open triangle) and hosting a BHS containing at least 10 BHs (filled points).The colour-coded map identifies the number of BHs in the BHS. The purple open diamonds represent the observed population of MW GCs \citep{harris10}. }
\label{comp}
\end{figure*}

With this purpose, we show in top panel of Figure \ref{comp} the central surface brightness $\Sigma$ and the average surface luminosity, defined above as $L/r_{\rm h,ob}^2$, for all the MOCCA models and for Milky Way GCs taken from the Harris catalogue \citep{harris96,harris10}.

Morever, in the bottom panel we compare the observational core radius ($r_{\rm c,ob}$) and half-light radius ($r_{\rm h,ob}$) of MOCCA models and actual GCs.

The $\Sigma - L/r_{\rm h,ob}^2$ plane is well divided in three different regions:

\begin{itemize}
\item one dominated by GCs with at most a few BHs after 12 Gyr;
\item one dominated by GCs containining an IMBH;
\item GCs containing a BHS.
\end{itemize}

BHS-dominated clusters have average surface luminosities in the range $10^2-10^{4.5}$ L$_\odot$ pc$^{-2}$ and surface brightness in between $10-10^4$ mag pc$^{-2}$. 
However, it is quite evident that the three regions are not uniquely defined and overlap at their boundaries. For instance, both 
systems hosting at least 10 BHs and those having up to 10 BHs gather in the same region of the plane. As expected, at a fixed value of the average surface luminosity, BHS dominate GCs having lower surface brightness. 

More interestingly, BHS are found in clusters having a $\Sigma$ value smaller than IMBHs. Conversely, IMBHs are grouped in a well defined region of the plane, with average surface luminosities above $3 \times 10^3$ L$_\odot$ pc$^{-2}$ and $\Sigma > 10^3$ mag pc$^{-2}$, although some models encroach upon smaller values.

Overlapping the distribution of MOCCA data with observed GCs from the Harris catalogue \citep[][2010 version]{harris10} shows immediately that a noticeable number of Galactic globulars may harbor an IMBH or a massive BHS, 
while in some others the BH population has been almost completely depleted due to high natal kicks and dynamical interactions.

Note that many Galactic GCs are expected to lie in the region characterized by ${\rm Log} ~L/r_{\rm h,ob}^2 = 2-4$. This preliminary comparison shows at a glance that many MW GCs can potentially host BHS characterised by relatively low-density and quite massive BHs, with average mass in between $14-22\Ms$.

Some interesting hints about the BHS-dominated GCs are also provided by the relation between the observational half-light and core radii.
Indeed, nearly all the MOCCA models hosting a BHS with more than 10 BHs have observational core radius larger than $\sim 0.3$ pc and half-light radius larger than 1 pc. Interestingly, observed GCs follow the same trend of our MOCCA subsample in the $r_{\rm h,ob}-r_{\rm c,ob}$ plane. Note that IMBH-dominated models gather in a small portion of the plane, being characterized by relatively low core radii $r_{\rm c,ob}\lesssim 1$ pc \footnote{
We note here that $r_{\rm c,ob}$ might be ill-defined in models with a central IMBH, because of the presence of a very massive object in the cluster centre. In our forthcoming work, mostly focused on IMBHs, we will compute this quantity carefully and compare with the preliminary estimates provided here.}, and nearly constant half-light radii $r_{\rm h,ob}\simeq 1-4$ pc.
It appears evident that models with a small content in BHs overlap to both IMBHs and BHS systems, making difficult to remove the degeneracy between all the three possibilities.

In a companion paper, we identified a sample of 29 Galactic globulars that may host a central BHS. Using the correlations presented in this paper, we calculated for all these clusters the BHS main properties (mass, size, average mass) and provide an estimate of the number of retained BHs, either single or as the component of a binary system \citep{askar18}.
Interestingly, a number of these targeted GCs are already known in literature for being host of several phenomena related to BH physics. 

One of the GCs falling in our selection is NGC 3201, which recently made the headlines thanks to the discovery of a BH in a detached binary \citep{giesers18}. As discussed in more detail in our companion paper, our estimate for this cluster are $\sim 20$ BHs as components of a binary system, and $\sim 10^2$ single BHs.

\section{Conclusions}
\label{sec:end}

In this paper, we used results from hundreds of GC models that were evolved as part of the MOCCA-Survey Database I to find correlations that can be used to infer the presence of a BHS in GCs using observational parameters.
Our main results can be summarized as follows:
\begin{itemize}
\item[i.] we provide a novel definition for a BHS and its boundaries in GC, according to which a BHS is the ensemble of BHs enclosed within the typical radius, $R_\bhs$, containing half the mass in BHs and half the mass in stars. The idea beyond $R_\bhs$ is conceptually similar to the definition of influence radius, i.e. the length scale over which a supermassive BH dominate the dynamics in a galactic nucleus;
\item[ii.] we define five main structural parameters for the BHS: radius, total mass, number of BHs, typical density, average mass and binarity. We found several correlations linking the quantities with each other. In particular, we found that the number of BHs increases smoothly with the total BHS mass, thus implying that heavier BHS are composed of heavier BHs, on average. At the same time, we found that heavier BHS have larger sizes and lower densities;
\item[iii.] comparing the BHS typical density and the host cluster central density, we found that at fixed GC density, heavier BHS are characterised by lower densities. On the other hand, in general the denser the GC the denser the BHS;
\item[iv.] the average mass of the BHs populating a BHS depends strongly on the BHS size and, intriguingly, it depends also on the average mass of the stars mixed within $R_\bhs$. In fact, higher star masses corresponds to lower BHs masses and vice-versa. In general, we found that heavier BHs are more likely to reside in the most extended BHS;
\item[v.] the BHS structural properties contains information about the dynamical age of the host cluster: high-density BHS, containing heavier BHs, reside in  dynamically old GCs, while heavier BHS, with much lower densities are found in dynamically young GCs with long relaxation times;
\item[vi.] the relation between the GC dynamical age and the BHS properties is due to the nature of relaxation process. If a GC is dynamically young, its population of heavy BHs did not undergo yet ejection in strong dynamical interactions. Since the heavier the BHs, the harder the binary in which they bind and consequently, the larger the energy budget that they can exchange with the environment thus heavy BHs lead to sparser and larger BHSs. On the other hand, in a dynamically old GC, the most massive BHs underwent segregation and core-collapse, with consequent formation and ejection of massive BHs and BH binaries. The resulting BHS will have lower mass and higher concentration;
\item[vii.] the BHS properties also reflect the number of binaries containing at least one BH in the GC. In particular, we found that larger BHSs correspond to lower number of binaries, normalized to the total BHs in the BHS itself. Moreover, the fraction of binaries is lower at larger BHS sizes. This can again be related to the GC dynamical status, as dynamically young GCs (large and heavy BHSs) will have experienced no or little binary formation involving single or double BHs;
\item[viii.] the BHS properties are inherited by the initial GC BHs population. Indeed, we found that the BHS mass is about $70-80\%$ of the total BH mass if its number of BHs is $N_\bhs \gtrsim 100$, while it is below $70\%$ for smaller values of $N_\bhs$;
\item[ix.] we found a tight correlation in what we call ``the fundamental plane for BHSs'', defined by the GC average surface luminosity $L/r_{\rm h,ob}^2$ and the BHS density. This, combined with the $\rho_\bhs - R_\bhs$ and the $R_\bhs -M_\bhs$ correlations allows us to fully characterize the BHS properties from two observational quantities;
\item[x.]  we found that BHS distribute in a well defined region of the plane delimited by the GC central surface brightness $\Sigma$ and its average surface luminosity $L/r_{\rm h,ob}^2$, quite detached by GCs with no BHs at 12 Gyr and from GCs hosting an IMBH. Comparing our models with observed GCs provided by the updated \cite{harris10} catalogue, we found that many Galactic GCs likely host a BHS with average masses in the range $14-22\Ms$;
\item[xi.]  we are also able to apply a similar procedure to MOCCA models hosting an IMBH, defining also in this case a typical radius $R_\ibh$ and a typical density $\rho_\ibh$ that is correlated with the IMBH mass and the GCs average surface luminosity density. A more detailed investigation will be carried out in a future paper. 
\end{itemize}
The approach presented here aims at providing a simple and rapid treatment that serves to select in an easy way potentially interesting candidates for detailed numerical studies or dedicated observations. Clearly, while we focused the attention on Galactic GCs, the relations described in the paper can be, in principle, used to identify BHS or IMBHs in extragalactic GCs.  

In our companion paper, we show how the information obtained through our simple scaling relation can be used to provide detailed information about the dark content of 29 Galactic GCs.

\section{Acknowledgement}
We would like to thank the anonymous referee whose comments and suggestions allowed us to significantly improve the contents of this manuscript. MAS acknowledges the Sonderforschungsbereich SFB 881 "The Milky Way System" (subproject Z2) of the German Research Foundation (DFG) for the financial support provided and the Nicolaus Copernicus Astronomical Center for the hospitality given during the development of this work. MG and AA were partially supported by the National Science Center (NCN), Poland, through the grant UMO-2016/23/B/ST9/02732. AA was also supported by NCN, Poland, through the grant UMO-2015/17/N/ST9/02573.

\clearpage
\footnotesize{
\bibliographystyle{mnras}
\bibliography{biblio}
}

\end{document}